\documentclass[twocolumn,english,aps,superscriptaddress, pra,groupedaddress]{revtex4-1}

\usepackage[latin9]{inputenc}
\usepackage{amsmath}
\usepackage{amssymb}
\usepackage{graphicx}
\usepackage{babel}
\usepackage{mathrsfs}
\usepackage{amsfonts}
\usepackage{epstopdf}
\usepackage{caption}
\usepackage{multirow}

\numberwithin{equation}{section}
\renewcommand\theequation{\arabic{section}.\arabic{equation}}

\begin{document}

\title{Non-Perturbative Dynamical Effects in Nearly Scale Invariant Systems: \\ The Action of Breaking Scale Invariance}

\author{Jeff Maki$^1$}
\author{Li-Ming Zhao$^{1,2}$}
\author{Fei Zhou$^{1}$}

\affiliation{1) Department of Physics and Astronomy, University of British Columbia, Vancouver V6T 1Z1, Canada \\ 2) Center for Theoretical Physics, Capital Normal University, Beijing, China}

\date{\today}

\begin{abstract} 

In this work we develop a general formalism that categorizes the action of broken scale invariance on the non-equilibrium dynamics of non-relativistic quantum systems. This approach is equally applicable to both strongly and weakly interacting systems. We show that any small deviation from the strongly interacting fixed point, in three spatial dimensions, leads to non-pertubative effects in the long time dynamics, dramatically altering the dynamics observed at the scale invariant fixed point. As a concrete example, we apply this approach to the non-equilibrium dynamics for the interacting two-body problem, and for a non-interacting quantum gas in the presence of an impurity, both in three spatial dimensions. Slightly away from the resonantly-interacting scale invariant fixed point, we show that the dynamics are altered by a non-perturbative log-periodic beat. The presence of the beat depends only on deviating from the resonant fixed point, while the frequency depends on the microscopic parameters of the system.

\end{abstract}

\maketitle

\section{Introduction}

Ever since the creation of atomic gases, the study of their non-equilibrium dynamics has been a rewarding pursuit. In recent years, the technological prowess displayed in experiments has allowed for  an unprecedented study in dynamics. The number of dynamical studies available cover a plethora of phenomena, ranging from quench dynamics 
\cite{Bosenova, Polkovnikov11, Resonance, Makotyn14}, thermalization and localization \cite{Polkovnikov11, Kaufman, Gring12, Schreiber15}, and periodic driving
\cite{Eckardt17, Meinert16, Jotzu14}. On the theoretical side, there has been considerable progress in understanding these dynamical phenomena \cite{Polkovnikov11, Eckardt17, Spivak04,Heyl13, Gurarie14,Pal10}, but often this is limited to numerical or semi-classical methods.


One fruitful approach for understanding such a complex problem is symmetry. The presence of symmetries in a system will undoubtedly lead to simplifications. One such symmetry that still peaks the interest of theorists and experimentalists alike is the symmetry associated with scale invariance. When a system possesses scale invariance, the governing equations of motion remain unchanged when subject to dilations of the spatial and temporal coordinates:

\begin{align}
 \vec{r}_i'  &= e^{-b}  \vec{r}_i   & i &=1,2,...,N   &  t' &= e^{-2b} t,
\label{scaling}     
\end{align}

\noindent where $\lbrace \vec{r}_i \rbrace$ denotes the position of the $N$ particles in the system with $i = 1,2,...N$, and $b$ is a scaling factor.

Thankfully, cold atoms offer a number of controllable scale invariant and nearly scale invariant systems; such as the two dimensional quantum gas 
\cite{Dalibard07,Chin11,Fenech16, Kohl12}, the three dimensional degenerate Fermi gas at unitarity 
\cite{Ohara02, Ku12, Cao11}, the three dimensional degenerate Bose gas at unitarity \cite{Makotyn14}, and the two-dimensional Fermi gas at a p-wave resonance \cite{Jiang17}. For these systems, scale invariance has been applied to obtain universal relations for quantities such as the equation of state, and bulk viscosity \cite{Nishida06, Jiang14, Jiang17}. The application of scale invariance in cold atoms is a strong motivation for a more systematic study of scale symmetry in non-equilibrium dynamics.

The primary method for theoretically understanding the dynamics of interacting cold atom systems has been the scaling solution \cite{Castin_Dum, Kagan96, Dalfovo97, Menotti02}. The scaling solution is a variational approach that is often applied to semiclassical approximations. Although the interacting quantum gas is generally not scale invariant, it was noticed that the scaling solution provides an approximate description of the dynamics \cite{brev}. An extension of this approach to examine the effects of broken scale invariance on the dynamics of a two dimensional Bose gas, with attractive interactions, was presented in Ref.~\cite{Maki16}.

One of the first attempts at understanding the role of scale invariance on fully quantum dynamics was presented in Ref.~\cite{Rosch97}. In this work, the motion of the moment of inertia was examined for an interacting two-dimensional Bose gas in a harmonic trap. This system was shown to have an approximate hidden SO(2,1) symmetry. This approximate symmetry is related to conformal symmetry up to the quantum anomaly \cite{Hofmann12,Olshanii10, Gao12}.  In terms of the moment of inertia, it was shown that the exact symmetry fixes the frequency of the breathing mode at twice the trap frequency. Even though it was first predicted for bosons, this approximate symmetry is present for two-dimensional Fermi gases, \cite{Rosch97,Taylor12}, and has been observed experimentally for both bosons and fermions \cite{Chevy02, Kohl12}. More recently, the role of scale invariance on the dynamics of quantum gases in time dependent harmonic traps was studied experimentally \cite{Deng16}.

A full description of scale invariant quantum dynamics can be provided by the general equations of motion. Equations of motion are useful for addressing the role of scale invariance on the dynamics of global observables \cite{Rosch97,Deng16, Blume16, Hofmann12,Olshanii10, Gao12}. However this approach becomes rather difficult, if not unsolvable, when the system is not fine tuned to scale invariance. In addition, this approach does not include much local space-time information that is contained in the wave function itself. A microscopic approach, which takes into account the SO(2,1) symmetry on the many body wave function, was put forward in Ref.~\cite{Castin}. In particular, this approach has been used to study both the energetics and dynamics of unitary Fermi gases in three dimensions \cite{Castin, Qu16}, of the Tonks gas in one dimension \cite{Minguzzi05}, and for harmonically trapped scale invariant systems \cite{Moroz12}.

For a given physical system, scale invariance is the result of the system residing at a fixed point of the renormalization group equations \cite{Wilson83,Sachdev}. Most systems will possess multiple fixed points, each characterized by their own infra-red stabilities and critical exponents. In the context of three dimensional cold gases, there are two fixed points available: the non-interacting and resonantly-interacting quantum gas. When a system resides at a fixed point, certain dynamics, following the Heisenberg equations of motion, will be dictated by the scale symmetry alone, and neither on the stability of the fixed point, nor the microscopic features of the system. 

This begs the question, how the non-equilibrium dynamics will be affected by the stability and universality of the fixed point, when one deviates from the fixed point. Will the non-equilibrium dynamics be completely equivalent near different fixed points, or will they fall into their own stability and universality class? To perform such a study one must extend the analysis of dynamics, not only to the resonant fixed point, but also to the vicinity of the fixed point, where the scale symmetry is broken explicitly, but slightly. Although there has been numerous theoretical studies on strongly interacting quantum gases using scale symmetry \cite{Nishida06, Zhang09, Jiang14,Tan08,Jiang16}, a complete understanding of three-dimensional resonant quantum gases has yet to be achieved. It is not surprising that the categorization of the dynamics near the different scale invariant fixed points has been uncharted waters for theorists.

In this article we lay down the general framework for categorizing the dynamics away from a given fixed point.  This formalism is valid for arbitrary number of particles, regardless of their statistics. We show that the dynamics do indeed depend upon whether one deviates from the resonantly-interacting or non-interacting fixed point. To perform this study, we employ the microscopic wave function which explicitly exhibits conformal symmetries. This approach is useful for studying the dynamics slightly away from a scale invariant point, as it expands the dynamics in terms of a complete set of wave functions whose unitary evolution is equivalent to a time dependent rescaling.

In this work we state two main results: i) for three dimensional systems away from the resonant fixed point, we find that the dynamics are fundamentally different in comparison to the dynamics at the resonant fixed point itself. Near resonance, the scale invariant dynamics are broken by a non-trivial logarithmic time dependence, the presence of which is non-pertubative, and depends only on deviating from the resonant fixed point. ii) We provide an experimental proposal on how to detect these predictions using an ensemble of two-body systems in an optical lattice.

As a concrete example we employ this formalism to study the dynamics of two systems:  the two-body problem with short ranged interactions, and the non-interacting quantum gas in the presence of a short ranged impurity, both in three spatial dimensions. In general, these two systems are not scale invariant, thanks to the presence of a finite scattering length, $a$. However, when $a$ is infinite or zero, these systems will become scale invariant.  In cold atom experiments both scale invariant fixed points can be achieved thanks to the presence of Feshbach resonances \cite{Resonance}.

The remainder of the article is organized as follows: in Section \ref{sec:castin} we review the many body wave function. The dynamics of scale invariant systems are studied in Section \ref{sec:SI}. The general formalism for describing the non-equilibrium dynamics away from a fixed point is then given in Sec. \ref{sec:dynamics_near_formalism}. The dynamics of the two body problem are then considered in Sec. \ref{sec:two_body}. In Sec. \ref{sec:near_res_imp} we consider the dynamics for the non-interacting quantum gas in the presence of an impurity.  We then conclude with a discussion of the results in Section \ref{sec:disc}.

\section{Dynamics in an Expanding Co-Moving Frame}
\label{sec:castin}

We begin by examining the time dependent many body Schrodinger equation for a system of $N$ particles with spin-independent interactions, $V(\vec{r})$, and in an external potential, $U(\vec{r})$, in three spatial dimensions:

\begin{align}
i &\partial_t \psi\left( \lbrace\vec{r}_i, \sigma_i \rbrace, t \right)= H  \psi\left( \lbrace\vec{r}_i, \sigma_i\rbrace, t\right), \nonumber \\
H &= \sum_i\left[ -\frac{1}{2} \nabla_i^2 + U(\vec{r}_i)  \right] +\frac{1}{2}\sum_{i,j} V(\vec{r}_i - \vec{r}_j),
\label{eq:Schrodinger} 
\end{align}

\noindent where the atomic mass, $m$, and $\hbar$ have been set to unity. $\vec{r}_i$ and $\sigma_i$ designate the position and spin of the $i$th particle, respectively.

For systems with scale invariance or that are nearly scale invariant, it is convenient to introduce the wave function:

\begin{eqnarray}
\psi(\lbrace\vec{r}_i, &&\sigma_i \rbrace, t) = \nonumber \\
&& \frac{1}{\lambda^{3N/2}(t)} e^{\frac{i}{2} \sum_ir_i^2 \dot{\lambda}(t)/\lambda(t)} \phi\left( \lbrace \frac{\vec{r}_i}{\lambda(t)},\sigma_i \rbrace, \tau(t) \right). \nonumber \\
\label{eq:Castin_wf}
\end{eqnarray}

\noindent In Eq.~(\ref{eq:Castin_wf}), $\lambda(t)$ is a time dependent scaling factor with units of length, and $\tau(t)$ is a dimensionless effective time. The wave function, $\phi(\lbrace \vec{r}_i / \lambda(t),\sigma_i \rbrace, \tau(t))$, contains all the many body information of the original wave function, and is both appropriately symmetrized and normalized. This wave function was first introduced in Ref. \cite{Castin}. The validity of this wave function for non-relativistic quantum systems was discussed in Ref.~\cite{Gritsev10}. 

Eq.~(\ref{eq:Castin_wf}) is motivated by the scale transformation given by Eq.~(\ref{scaling}). It combines a time dependent rescaling of the spatial coordinates with a gauge transformation. The effect of this combination is to separate the simple rescaling dynamics governed by $\lambda(t)$ from the non-trivial dynamics governed by $\tau(t)$.


After substituting Eq.~(\ref{eq:Castin_wf}) into Eq.~(\ref{eq:Schrodinger}), it is possible to find a Schrodinger equation that describes the non-trivial dynamics contained in the field $\phi(\lbrace r_i / \lambda(t), \sigma_i \rbrace, \tau(t))$:

\begin{align}
& i \frac{\partial}{\partial \tau} \phi(\lbrace \vec{x}_i, \sigma_i \rbrace, \tau) = \tilde{H}(\tau) \phi(\lbrace \vec{x}_i, \sigma_i \rbrace, \tau), \nonumber \\
& \tilde{H}(\tau) = \sum_i\left[-\frac{1}{2}\tilde{\nabla}_i^2 + \frac{x_i^2}{2} + \lambda^2(\tau) U(\lambda(\tau) \vec{x}_i)\right] \nonumber \\
&+ \frac{1}{2}\sum_{i,j}\lambda^2(\tau) V(\lambda(\tau)(\vec{x}_i - \vec{x}_j)).
\label{eq:nirf_se}
\end{align}

\noindent In Eq.~(\ref{eq:nirf_se}), the effective spatial and temporal coordinates in Eq.~(\ref{eq:nirf_se}) are defined as:

\begin{align}
\vec{x}_i &= \vec{r}_i /\lambda(t), & i &=1,2,...,N \nonumber \\
\tau(t) &= \arctan\left( \frac{t}{\lambda_0^2} \right), & 0& \leq \tau < \pi/2 \nonumber \\
\lambda(t) &= \lambda_0 \sqrt{1 + \frac{t^2}{\lambda_0^4}},& \lambda(\tau)& = \lambda_0 \sec(\tau) .
\label{eq:coordinates}
\end{align}

\noindent The full derivation of Eqs.~(\ref{eq:nirf_se}) and (\ref{eq:coordinates}) is given in Appendix ~\ref{app:comoving_frame}.

Eq.~(\ref{eq:nirf_se}) is just the Schrodinger equation for an interacting quantum gas in a harmonic trap. Physically, this transformation is equivalent to studying the dynamics in an expanding co-moving frame with coordinates given by Eq.~(\ref{eq:coordinates}). The harmonic potential in Eq.~(\ref{eq:nirf_se}) is simply a fictitious force which appears because one is working in this co-moving frame. The penalty for using this transformation is that the interactions and external potentials now acquire time dependence.

\section{Dynamics at a Scale Invariant Fixed Point}
\label{sec:SI}
When a quantum system in the laboratory frame possesses scale invariance, the Schrodinger equation, Eq.~(\ref{eq:Schrodinger}), is invariant under Eq.~(\ref{scaling}). For this to be true, a scale invariant Hamiltonian in the laboratory frame, $H_s$, must rescale like the time derivative: $H_s' = e^{2b} H_s$. This implies that for a system to be scale invariant, any potential must scale like the time derivative. That is, a scale invariant external potential, $U_s(\vec{r})$, and interaction potential, $V_s(\vec{r})$, must satisfy: $U_s( \vec{r}e^{-b}) = e^{2b} U_s(\vec{r})$, $V_s(\vec{r}e^{-b}) = e^{2b} V_s(\vec{r})$,
\ for some scaling factor, $b$. The main consequence of the scale symmetry on Eq.~(\ref{eq:nirf_se}), is that any scale invariant Hamiltonian in the laboratory frame,  will transform into a time independent Hamiltonian in the co-moving reference frame, $\tilde{H}$:

\begin{equation}
\tilde{H} = \sum_i \left[ -\frac{1}{2}\tilde{\nabla}^2_i + \frac{x_i^2}{2} + U_s(\vec{x}_i)\right] + \frac{1}{2} \sum_{i,j} V_s(\vec{x}_i-\vec{x}_j).
\label{eq:hs}
\end{equation}

Although $H_s$ in the laboratory frame is scale symmetric, $\tilde{H}$, defined in the co-moving frame, no longer possesses scale symmetry. However,  $\tilde{H}$ does form part of the conformal, or so(2,1), algebras \cite{Hagen72, Niederer72,Castin, Nishida07}. One can show that this symmetry guarantees that the spectrum of Eq.~(\ref{eq:hs}) contains a series of evenly spaced states: $E_n = 2n +E_0$,  $n = 0,1,2,...$. This spectrum is known as the confromal tower. In general, there can be multiple conformal towers present, each with a different ground state eigenvalue, $E_0$. This ground state energy is related to the scaling dimension of a given primary operator \cite{Nishida07}. In terms of dynamics, the breathing modes in two-dimensional quantum gases are an approximate conformal tower, a result of the approximate SO(2,1) symmetry \cite{Rosch97, Taylor12, Chevy02, Kohl12}.

In the laboratory frame, the conformal tower states of Eq.~(\ref{eq:hs}) are given by \cite{Note}:

\begin{eqnarray}
\psi_n(\lbrace\vec{r}_i \rbrace&&, t) = \nonumber \\
&& \frac{1}{\lambda^{3N/2}(t)} e^{\frac{i}{2} \sum_ir_i^2 \dot{\lambda}(t)/\lambda(t)} \phi_n\left( \lbrace \frac{\vec{r}_i}{\lambda(t)}\rbrace \right) e^{-i E_n \tau(t)}. \nonumber \\
\label{eq:Castin_solution}
\end{eqnarray}

\noindent Although the confromal tower states are not eigenstates of the original Hamiltonian, Eq.~(\ref{eq:Schrodinger}), they are nonetheless characterized by a simple global phase {\em as if they were standard eigenstates}:

\begin{align}
\Gamma_n(t) &= E_n \tau(t) = \int_0^t dt' \ \frac{E_n}{\lambda^2(t')}, \nonumber \\
\Gamma_n(t\gg \lambda_0^2) &\approx E_n(\pi/2+ O(\lambda_0^2 / t)).
\label{eq:phase}
\end{align}

As these phases appear in an identical way as the phases of standard eigenstates, it is tempting to introduce the concept of quasi energies associated with $\Gamma_n(t)$:

\begin{align}
E_n^{quasi} &= \partial_t \Gamma_n(t)=E_n / \lambda^2(t).
\label{eq:quasi_energy}
\end{align}

\noindent These quasi energies turn out to be related to the physical energy of a conformal state in the laboratory frame. By using the so(2,1) algebra, one can show that:

\begin{equation}
\langle n |H_s| n \rangle = \frac{1}{2 \lambda_0^2} E_n,
\end{equation}

\noindent where $\langle \lbrace \vec{x}_i \rbrace| n \rangle = \phi_{n}(\lbrace \vec{x}_i \rbrace)$. For these reasons, the quasi energies defined here turn out to be the most useful to our understanding of scale invariant dynamics and the relevance of the perturbations breaking scale invariance.

In the long time limit, $t \gg \lambda_0^2$, the quasi-energies in the laboratory frame will be compressed as $t^{-2}$. In the co-moving frame, these conformal tower states states do not evolve with time, and the spectrum is rigid. The differences between the co-moving and laboratory frames are shown in Fig.~(\ref{fig:lab_v_conformal}). In either case, the phase factor will approach a constant at long times.

Since the dynamical phase approaches a constant, see Eq.~(\ref{eq:phase}), the wave function for a scale invariant system, in the co-moving frame, freezes. When this happens all the long time dynamics in the laboratory frame will be related by a time-dependent rescaling. To be explicit, if one considers a local operator, $O$, of scaling dimension, $d_O$, one obtains the approximate scaling solution for the dynamics:

\begin{figure}
\includegraphics[scale=0.45]{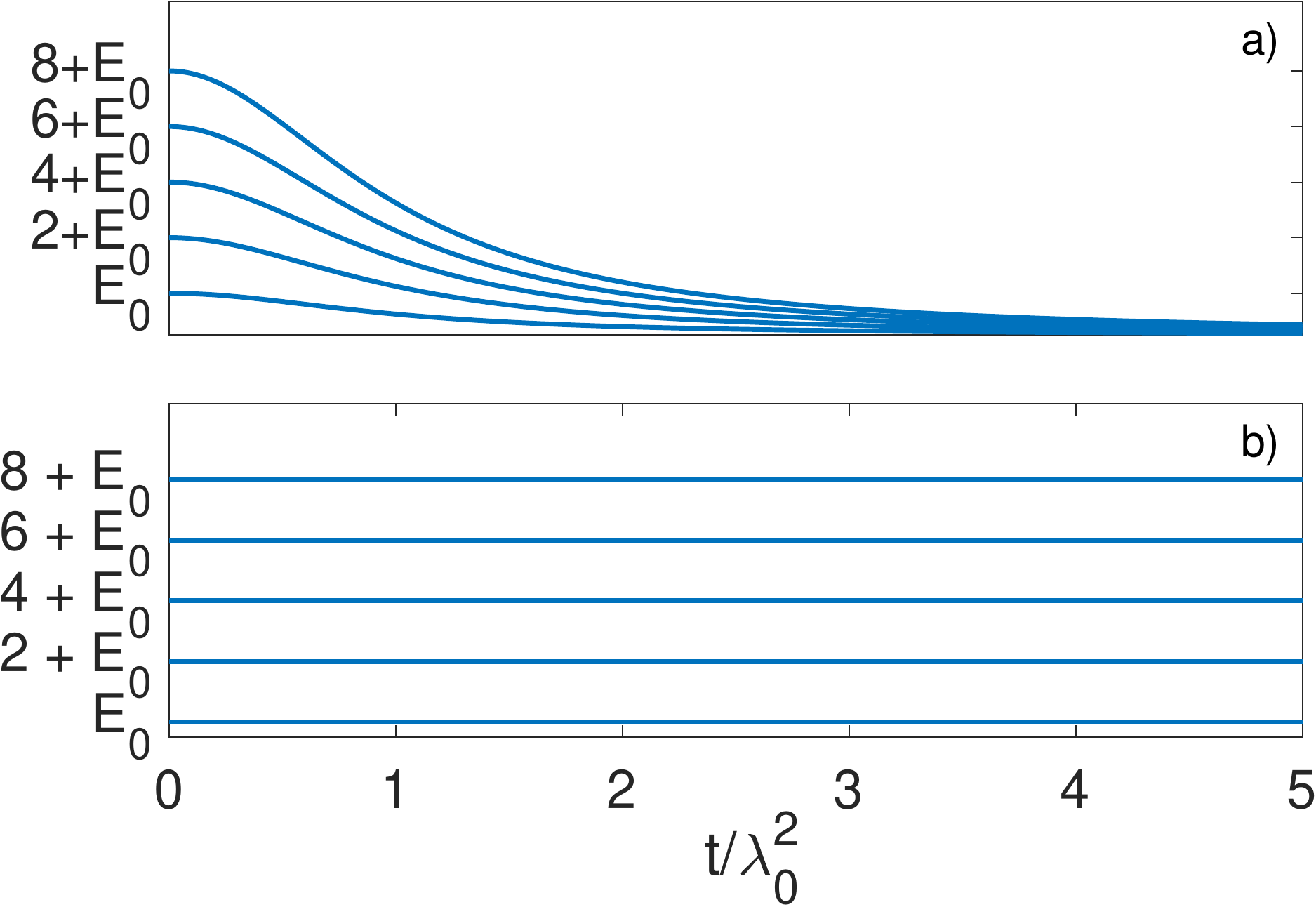}
\caption{The conformal tower states in both a) the laboratory frame, and b) in the co-moving frame. In the laboratory frame, the eigenstates of Eq.~(\ref{eq:hs}), are evenly spaced but contract like $~t^{-2}$ in the long time limit, $t \gg \lambda_0^2$, see Eq.~(\ref{eq:quasi_energy}). In the co-moving frame the spectrum does not evolve with time.}
\label{fig:lab_v_conformal}
\end{figure}

\begin{align}
\lim_{t \rightarrow \infty}\langle O \rangle(t) &=  \int \frac{d^{3N}\lbrace \vec{r}_1 \rbrace}{\lambda^{3N}(t)} \left|\phi\left(\lbrace \frac{\vec{r}_i}{\lambda(t)}\rbrace,\tau(t)\right)\right|^2 O(\lbrace \vec{r}_i\rbrace), \nonumber \\ 
&\approx \frac{1}{\lambda(t)^{d_O}}\int d^{3N}\lbrace \vec{x}_i \rbrace \ |\phi(\lbrace\vec{x}_i\rbrace, \pi/2)|^2  \rbrace) \tilde{O}(\lbrace\vec{x}_i \rbrace), \nonumber \\
&\approx \frac{1}{\lambda^{d_O}(t)} \langle \tilde{O}\rangle (\pi/2),
\label{eq:si_dynamics_physical_observable}
\end{align}

\noindent where $\langle \tilde{O}\rangle (\pi/2)$ is the asymptotic expectation value of the operator in the co-moving frame. Eq.~(\ref{eq:si_dynamics_physical_observable}) only depends on the scale symmetry of the initial Hamiltonian in the laboratory frame. It does not depend on the statistics or the number of particles in the system and applies to general scale invariant many body systems.

To give a more concrete example, consider the moment of inertia operator, $r^2$ of an $N$-particle system:

\begin{align}
r^2  &\equiv \frac{1}{N} \sum_{i=1}^N r_i^2 \nonumber \\
\lim_{t \rightarrow \infty}\langle r^2 \rangle(t) &= \lambda^2(t) \int d^{3N} \lbrace \vec{x}_i \rbrace  |\phi(\lbrace\vec{x}\rbrace_i, \pi/2)|^2 x_1^2,\nonumber \\
&= \lambda^2(t) \langle x^2 \rangle (\pi/2),
\label{eq:si_dynamics_rsquared}
\end{align}

\noindent where $\langle x^2 \rangle(\tau)$ is the moment of inertia for the $N$-particle system in the co-moving reference frame. 

As we will see in the next section, the dynamical phase is equally important for understanding the dynamics away from a scale invariant fixed point.

\section{Dynamics away from a Scale Invariant Fixed Point}
\label{sec:dynamics_near_formalism}

In this section we put forward a general formalism for studying the dynamics away from a scale invariant fixed point. This formalism allows one to contrast the dynamics near different scale invariant fixed points. This discussion does not depend on the number of particles or their statistics, only on deviating from a given fixed point.

We first write the effective Hamiltonian in Eq.~(\ref{eq:nirf_se}), $\tilde{H}(\tau)$, as a sum of two parts:

\begin{equation}
\tilde{H}(\tau) = \tilde{H} + \delta \tilde{H}(\tau),
\end{equation}

\noindent where $\tilde{H}$ is the scale invariant Hamiltonian in the laboratory frame, transformed into the co-moving frame, Eq.~(\ref{eq:hs}), and $\delta \tilde{H}(\tau)$ is the deviation from $\tilde{H}$. As discussed in the previous section, $\tilde{H}$ is time independent due to the scale symmetry in the laboratory frame, and thus all the time dependence of the interaction is contained in $\delta \tilde{H}(\tau)$.

For the most general deviation that we want to consider, the deviation operator in the co-moving frame can be written in the form:

\begin{equation}
\delta \tilde{H}(\tau) = \tilde{h} \left(\frac{\lambda(\tau)}{a}\right)^{\alpha}= \tilde{h} \left(\frac{\lambda_0}{a}\right)^{\alpha}\sec^{\alpha}(\tau).
\label{eq:deviation_scaling}
\end{equation} 

The exponent, $\alpha$, in Eq.~(\ref{eq:deviation_scaling}) is fully controlled by the scaling behaviour of the Hamiltonian {\em near} the scale invariant fixed point from which the system is deviating. That is, $\alpha$ only depends on the renormalization flow near the fixed point, and hence the universality class of the fixed point. Although the dynamics discussed in Sec.~\ref{sec:SI} only rely on the presence of scale invariance, disregarding the types of fixed points involved, the quantum dynamics when deviating from a scale invariant fixed point strongly depends on which fixed point is involved. To highlight this point, we further look into the scaling behaviour of the short range interaction, or the renormalization of the interaction constant, $g$.

In three dimensional (s-wave) scattering, the two scale invariant fixed points correspond to i) a non-interacting gas and ii) a resonant gas. The scaling behaviour near these two special models can be conveniently studied via the renormalization group equations for the dimensionless coupling constant, $\tilde{g} = g \Lambda/2\pi^2$. Under a scale transformation, where the ultra-violet cut-off transforms as $\Lambda ' \rightarrow \Lambda e^b$, $\tilde{g}(b)$ follows the differential equation:

\begin{equation}
\frac{d\tilde{g}}{d b}=\beta(\tilde{g})=\tilde{g}+\tilde{g}^2.
\label{eq:RG}
\end{equation}

Eq.~(\ref{eq:RG}) describes how the interaction constant $\tilde{g}(b)$, or the interaction Hamiltonian, rescales as a function of $e^b$, or  equivalently, the rescaled momentum cut-off. It contains two scale invariant fixed points, which correspond to the two zeroes of the beta-function, $\beta(\tilde{g})$. One is $\tilde{g}^*=0$,  the non-interacting theory, and the other is $\tilde{g}^*=-1$, the resonance fixed point with strong interactions. At these two points, $\tilde{g}(b)$ remains constant under the scale transformation, and is hence scale invariant.

Near the non-interacting fixed point, one can easily verify, by linearizing the beta function, that:

\begin{eqnarray}
\tilde{g} \sim \Lambda e^b,
\end{eqnarray}

\noindent as we take the infra-red limit, or as $b$ approaches $-\infty$.

This simply indicates the stability of the three dimensional free particle fixed point in the long wavelength limit. This analysis suggests that $\tilde{g}$ itself shall have the same scale dimension as the momentum $\Lambda e^b$, or equivalently the inverse of the length scale in the problem, $\lambda(t)$.
Hence $\alpha= -1$ in $\delta \tilde{H}(\tau)$.

However, near the resonant fixed point, $\tilde{g}^*$, the perturbation away from the fixed point rescales as:

\begin{eqnarray}
\delta\tilde{g}=\tilde{g}-\tilde{g}^* \sim \Lambda e^{-b},
\end{eqnarray}

\noindent as anticipated since the resonance fixed point is unstable in the infra-red limit, or when $b$ tends to minus infinity. This implies that $\delta \tilde{g}$, as well as $\delta \tilde{H}$, shall rescale as the length scale, 
$\lambda(t)$, in contrast to the free particle fixed point where $\delta \tilde{H}(\tau)$ is proportional to $1/\lambda(t)$. Consequently,  $\alpha=+1$ near resonance. In Appendix \ref{app:alpha}, we show this is indeed the case.
In general, one can show that the scaling exponent for a deviation from a given fixed point, $-\alpha$,  is exactly equal to the derivative of the beta-function at the fixed point:

\begin{eqnarray}
\alpha(\tilde{g}^*)=-\left. \frac{\partial \beta(\tilde{g})}{\partial \tilde{g}}\right|_{\tilde{g}=\tilde{g}^*}.
\end{eqnarray}

Note that since all the scaling analysis has been carried out without referring to the details of the energetics, but only on the universal aspect of the scaling properties, they are applicable to arbitrary number of particles,  $N$. This approach is valid in both the many-body limit when $N$ taken to be infinity and few body cases when $N=2,3,4...$.

Furthermore, after proper regularization, $\tilde{h}$ is a universal, time-independent, dimensionless, operator that only depends on the number of particles, $N$. In Appendix \ref{app:derivation}, explicit calculations for $\tilde{h}$ can be found for the deviation near the resonant, and non-interacting fixed points, for both the two-body problem, and for the non-interacting quantum gas in the presence of an impurity.

To solve the time dependent Schrodinger equation away from a given fixed point in the co-moving reference frame, Eq.~(\ref{eq:nirf_se}), it is convenient to use the interaction picture with respect to $\tilde{H}$. The formal solution to Eq.~(\ref{eq:nirf_se}) is given by \cite{Sakurai}:

\begin{eqnarray}
\phi(\lbrace \vec{x}_i \rbrace, \tau) &=& \sum_n C_n(\tau)e^{-i E_n \tau} \phi_{n}(\lbrace \vec{x}_i \rbrace), \nonumber  \\
C_n(\tau) &=& \langle n | T e^{-i \_0^{\tau} d\tau' \delta \tilde{H}_I(\tau')} |\psi_0 \rangle, \nonumber \\
\delta \tilde{H}_I(\tau) &=& e^{i \tilde{H}\tau}\delta \tilde{H}(\tau) e^{-i \tilde{H} \tau},
\label{eq:Int_Picture_definitions}
\end{eqnarray}

\noindent where $\langle \lbrace \vec{x}_i \rbrace| n \rangle = \phi_{n}(\lbrace \vec{x}_i \rbrace)$ are the eigenstates of Eq.~(\ref{eq:hs}) with energies $E_n=2n+E_0$,  $\langle \lbrace \vec{r}_i \rbrace | \psi_0 \rangle = \lambda_0^{-3N/2}\phi(\lbrace \vec{r}_i / \lambda_0 \rbrace)$ is the initial wave function, and $T$ represents the time ordering operator.


Although Eq.~(\ref{eq:Int_Picture_definitions}) gives the exact solution, it is instructive to first consider how the conformal tower states are hybridized by the interaction, as a function of time. For a given conformal tower state, the number of other states which are hybridized by the interaction, to lowest order, can be easily measured by the ratio of the strength of the interaction to the level spacing in the tower. In the laboratory frame the strength of the deviation will be given by $\delta H(t) \propto \lambda^{\alpha-2}(t)/a^{\alpha}$, while the quasi-energies are given in Eq.~(\ref{eq:quasi_energy}). As a result the number of states effectively coupled to a given state by the lowest order interaction effect is given by:

\begin{align}
N_{coupled} &= \frac{1}{a^{\alpha}}\frac{\lambda^{\alpha-2}(t)}{2/\lambda^2(t)}\nonumber \\
&=\frac{1}{2}\left(\frac{\lambda(t)}{a}\right)^{\alpha} \approx \frac{1}{2}\left( \frac{t}{\lambda_0 a}\right)^{\alpha}
\label{eq:n_coupled}
\end{align}

\begin{figure}
\includegraphics[scale=0.45]{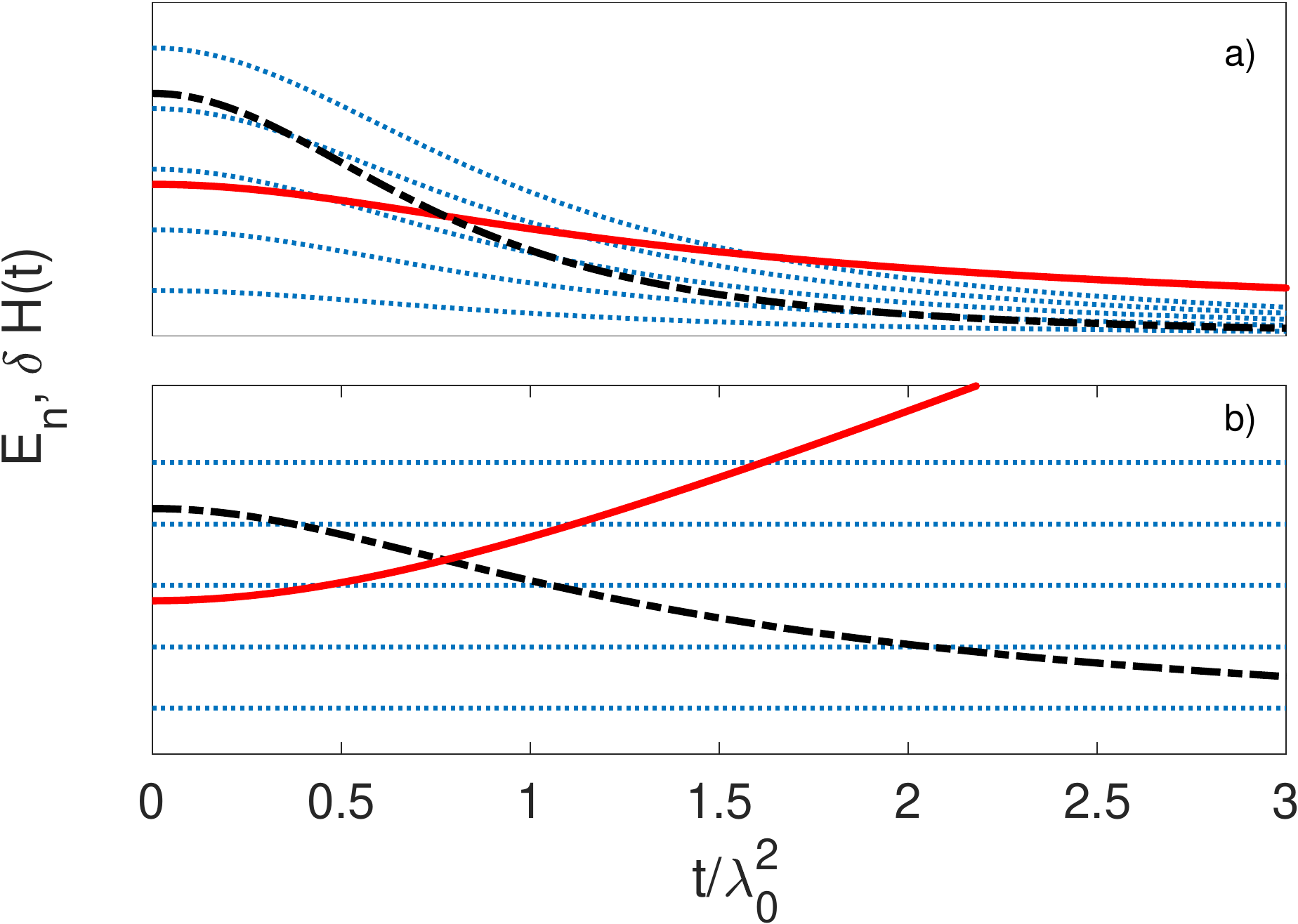}
\caption{Here we show the conformal tower spectrum and perturbation in both a) the laboratory frame, and b) the co-moving frame. The blue (dotted) lines correspond to the conformal tower states. The red (solid) and black (dash-dotted) lines correspond to deviations with scaling, $\alpha = 1$ and $\alpha = -1$, respectively. For scaling $\alpha = 1$, the interaction eventually couples more and more states, see Eq.~(\ref{eq:n_coupled}), with $N_{coupled} \propto t$. For $\alpha = -1$, the interaction vanishes with time, i.e. for long times, fewer and fewer states are coupled together with, $N_{coupled} \propto  1/ t$. We therefore expect a breakdown of time-dependent perturbation theory for $\alpha \geq 1$.}
\label{fig:spectrum_v_pert}
\end{figure}

\noindent For $\alpha \geq 1$, $N_{coupled}$ diverges as a function of time, i.e. more and more states will be affected by the perturbation. This indicates the potential breakdown of perturbation theory, as will be further proved later on. For $\alpha<1$, the perturbation grows either too slowly, or decreases too quickly, compared to the compression of the conformal tower states. In the co-moving frame, the energies do not evolve with time, but the strength of the perturbations on the other hand scales as $\lambda^{\alpha}(t)$, again, leading to the same result in Eq.~(\ref{eq:n_coupled}). For the cases $\alpha = 1$ and $\alpha = -1$, we compare the perturbation strength to the quasi-energies in Fig.~(\ref{fig:spectrum_v_pert}).

The divergence or disappearance of $N_{coupled}$ with time is reflected in the time-dependent perturbation theory. Consider the dynamics in the co-moving frame for simplicity. For $\alpha \geq 1$, the interaction diverges as: $(\pi/2 - \tau)^{-\alpha}$, in the long time limit, i.e. when $t \gg \lambda_0^2$, or equivalently, when $\tau$ approaches $\pi/2$. This divergence is linked to the growth of $N_{coupled} \propto t^{\alpha}$. To clearly see this, consider the expansion coefficients at long times, to first order in perturbation theory, for $\alpha = 1$:

\begin{align}
C_n(\tau)&\approx C_n(0) - i \frac{\lambda_0}{a} \log\left(\frac{1}{\pi/2 - \tau} \right) \nonumber \\
&\cdot \langle n | e^{-i \tilde{H} \pi/2} \tilde{h} e^{-i \tilde{H} \pi/2}| \psi_0 \rangle +...
\label{eq:long_time_pt}
\end{align}

\noindent As seen in Eq.~(\ref{eq:long_time_pt}), in the long time limit the strength of the interaction diverges logarithmically, regardless of how small the initial deviation is. For even larger scaling dimensions, $\alpha >1$, the divergence becomes more severe. For any scaling, $\alpha \geq 1$, perturbation theory will eventually become inadequate at describing the long time dynamics.

To illustrate this point further, consider the result of time dependent perturbation theory on the dynamics of the effective moment of inertia, $\langle x^2 \rangle(\tau)$, when $\alpha = 1$. To second order in the interaction, the moment of inertia will have the form:

\begin{align}
\langle x^2 \rangle (\tau) &= A + B \frac{\lambda_0}{a} \log\left( \frac{1}{\pi/2 - \tau}\right)  \sin(2 \tau) \nonumber \\
&+ D \frac{\lambda_0^2}{a^2} \log^2\left( \frac{1}{\pi/2 - \tau}\right) +....
\label{eq:pt_xs}
\end{align}

In Eq.~(\ref{eq:pt_xs}) the coefficients $A$, $B$, and $D$  depend on the zeroth, first, and second order expansions of Eq.~(\ref{eq:Int_Picture_definitions}), in terms of the interaction, respectively. The term proportional to $(\lambda_0/a)^0$ is the scale invariant result given in Eq.~(\ref{eq:si_dynamics_rsquared}). The term 
linear in $\lambda_0 /a$ is only important at short times because $\sin(2 \tau)$ vanishes as $\lambda_0^2 /t$, in the long time limit. Therefore at sufficiently long times the term linear in $\lambda_0/a$ will disappear while the quadratic term diverges. For this reason, it is necessary to develop a non-perturbative treatment for the breaking of scale invariance.

As shown in Appendix \ref{app:time_ordering}, it is possible to explicitly determine the leading long time behaviour of the time-ordering operator in Eq.~(\ref{eq:Int_Picture_definitions}), when $\alpha \geq1$:

\begin{align}
\lim_{\tau \rightarrow \pi/2} T &e^{-i \int_0^{\tau} d\tau' \delta \tilde{H}_I(\tau')} \approx  \nonumber \\
&\exp\left(i \frac{\lambda_0}{a} \tilde{V}_I\log(\pi/2 -\tau) \right), & \alpha &=1, \nonumber \\
& \exp\left(-i \tilde{V}_I\left(\frac{\lambda_0}{a}\right)^{\alpha}\frac{1}{\alpha-1} \frac{1}{(\pi/2 -\tau)^{\alpha-1}} \right), & \alpha &>1,
\label{eq:T_form}
\end{align}

\noindent where the operator, $\tilde{V}_I$, is the many body deviation from scale invariance in the interaction picture, evaluated at $\tau = \pi/2$:

\begin{equation}
\tilde{V}_I = e^{i \tilde{H} \pi/2} \tilde{h} e^{-i \tilde{H} \pi/2}.
\label{eq:V_matrix}
\end{equation}

\noindent This matrix is universal, time-independent and dimensionless. The concrete form of $\tilde{V}_I$ only depends on the number of particles involved, $N$.

The solution to the unitary dynamics in the co-moving frame is encapsulated in Eqs.~(\ref{eq:Int_Picture_definitions}), (\ref{eq:T_form}), and (\ref{eq:V_matrix}). This solution is equivalent to a time-independent problem with Hamiltonian, $\tilde{V}_I$, and time coordinate, $\int_0^t dt' N_{coupled}(t')/\lambda^2(t')$. In the co-moving frame, the long time dynamics will exhibit oscillations at frequencies determined by the spectrum of $\tilde{V}_I$. In the laboratory frame, this implies that there are non-trivial dynamics in addition to the trivial rescaling dynamics. Specifically, the dynamics of any local operator, $O(\lbrace\vec{r}_i \rbrace)$, with scaling dimension, $d_O$, will be given by:

\begin{align}
\lim_{t \rightarrow \infty} \langle O \rangle (t) &\approx \frac{1}{\lambda(t)^{d_O}} F\left( \left(\frac{\lambda_0}{a}\right) \log(t/\lambda_0^2) \right), & \alpha = 1, \nonumber \\
&\approx \frac{1}{\lambda(t)^{d_O}} F\left( \left(\frac{\lambda_0}{a}\right)^{\alpha} \frac{1}{\alpha-1} (t/\lambda_0^2)^{\alpha-1} \right), & \alpha > 1, 
\label{eq:physical_operator_scaling}
\end{align}

\noindent where $F(t)$ is a dimensionless function. Although the specific form of $F(t)$ depends on the number of particles, the specific observable, and the spectrum of the matrix, $\tilde{V}_I$, here we stress that the argument of this function is set only by the scaling of the deviation operator, $\alpha$. Therefore Eq.~(\ref{eq:physical_operator_scaling}), applies to general many body systems. In comparison to the result for scale invariant systems, Eq.~(\ref{eq:si_dynamics_physical_observable}), the long time dynamics are no longer equivalent to a simple time dependent rescaling.

A similar analysis is valid when $\alpha  <1$. In this case,  $N_{coupled}$ grows too slowly, or decreases in the time domain relative to the compression of the spectrum.  In this case, an application of time-dependent perturbation theory shows that the deviations from scale invariance will vanish in the long time limit. The dynamics in the co-moving frame will then quickly freeze, and the dynamics in the laboratory frame will become a simple time dependent rescaling. The only difference is that the profile of the wave function will be altered by the presence of interactions. The effect of the interactions on the expansion coefficients is well described by time dependent perturbation theory.

Eqs.~(\ref{eq:T_form}), (\ref{eq:V_matrix}), and (\ref{eq:physical_operator_scaling}) are the main theoretical results of this paper. These results are valid for a large number of non-relativistic quantum system close to a scale invariant fixed point. This formalism does not depend on the number of particles, nor their statistics, only on the scaling of the deviation, $\alpha$. This allows for a scaling analysis of the dynamics slightly away from a scale invariant fixed point, even for many body systems.

The remainder of this work will apply this formalism to two simple systems, the two-body problem, and the non-interacting quantum gas in the presence of an impurity. For these systems, the deviation occurs from fine tuning the potentials away from a scale invariant fixed point. For such systems, the deviation can be written in the form:

\begin{align}
\delta \tilde{H}(\tau) &= \frac{1}{2}\sum_{i,j}\lambda^2(\tau) V(\lambda(\tau)(\vec{x}_i - \vec{x}_j)) - V_s(\vec{x}_i-\vec{x}_j) \nonumber \\
&+ \sum_i \lambda^2(\tau) U(\lambda(\tau)\vec{x}_i) - U_s(\vec{x}_i).
\label{eq:delta_H}
\end{align}

\section{Dynamics of a Two-Body System}
\label{sec:two_body}

In this section we study the dynamics of two particles with short ranged s-wave interactions. The two particles can either be two bosons, two fermions in the spin singlet channel, or two distinguishable particles. For concreteness we consider two particles with identical masses. The Hamiltonian in the co-moving frame is given by Eq.~(\ref{eq:nirf_se}), with the external potential, $U(\vec{x}_i)$, set to zero.

In this analysis we consider a spherically symmetric, potential with a finite range, $r_0 \ll \lambda_0$. Since angular momentum is a good quantum number for such a potential, the scattering in different angular momentum channels will be uncoupled. At low energies, only the zero angular momentum scattering will be appreciable. The s-wave scattering will introduce a new length scale, the scattering length, $a$. This length scale will be present in the dynamics of the system, and breaks the scale invariance explicitly. However, the symmetry is restored when the system is non-interacting, $a=0$, or at resonance, $a=\infty$.

To facilitate our study, we also consider a wave function which is a superposition of only s-wave states with initially real expansion coefficients. This assumption is made to clearly highlight the departure from scale invariance on the dynamics near the resonant and non-interacting fixed points.

For two particles in the co-moving frame one can separate the center of mass coordinates, $\vec{X}$, from the relative ones, $\vec{x}$. The Schrodinger equation for the center of mass and relative motion, respectively, are:

\begin{align}
i \frac{\partial}{\partial \tau} \Phi(\vec{X},\tau) &= \left(-\frac{1}{4}\tilde{\nabla}_X^2 + 2 X^2 \right) \Phi(\vec{X},\tau), \nonumber \\
i \frac{\partial}{\partial \tau} \phi(\vec{x},\tau) &= \left(-\tilde{\nabla}^2 +\frac{1}{4} x^2 + \lambda^2(\tau)V(\lambda(\tau) \vec{x})\right) \phi(\vec{x},\tau),
\label{eq:relative}
\end{align}

\noindent where $\Phi(\vec{X},\tau)$ and $\phi(\vec{x},\tau)$ are the center of mass and relative wave functions, respectively. 

Eq.~(\ref{eq:relative}) is nothing more than two uncoupled single particle Schrodinger equations in an external potential. The dynamics for the center of mass will be equivalent to the trivial scale invariant dynamics, but the relative motions can be tuned away from scale invariance by means of the scattering length. Unless otherwise stated we will assume the  center of mass motion  is not entangled with the relative motion, and will only contribute trivially to the overall dynamics. The generalization to entangled motion is straightforward.

\subsection{Near Resonance}
For large scattering lengths, we expand the relative Hamiltonian around the resonant fixed point. As was discussed in Sec.~\ref{sec:dynamics_near_formalism}, also see Appendices~\ref{app:alpha} and \ref{app:derivation}, the scaling for the deviation is $\alpha = +1$. This means that the interaction can not be ignored in the long time limit and we need the non-perturbative formalism developed in Sec.~\ref{sec:dynamics_near_formalism}. Applying the formalism one can derive an analytical expression for the matrix, $\tilde{V}_I$, defined in Eq.~(\ref{eq:V_matrix}):

\begin{align}
\langle n | \tilde{V}_I | m \rangle &= f_n f_m & f_n &= \frac{\sqrt{2}}{\pi^{1/4}}\frac{(2n-1)!!}{\sqrt{(2n)!}}.
\label{eq:V_tilde}
\end{align}

\noindent The spectrum of $\tilde{V}_I$ can be determined by noting that each matrix element of $\tilde{V}_I$ is a product of two factors. For these separable matrices, there is a single non-zero eigenvalue, $v$, with eigenstate, $|v\rangle$:

\begin{align}
v &= Tr\tilde{V}_I = \sum_{n=0}^{n_{max}} f_n^2 & \langle n |v\rangle &= \frac{f_n}{\sqrt{v}}.
\label{eq:v}
\end{align}

\noindent The eigenvalue, $v$, diverges with the harmonic quantum number as: $\sqrt{n}$. We note that this sum is controlled by the energy scale set by the range of the potential: $n_{max} = r_0^{-2}\lambda_0^{2}$, where $r_0$ is the range of the potential with $r_0 \ll a$.

Inserting a complete set of eigenstates for $\tilde{V}_I$ allows one to solve Eq.~(\ref{eq:Int_Picture_definitions}) for the relative motion:

\begin{eqnarray}
\lim_{\tau \rightarrow \pi/2} & &\phi(\lbrace \vec{x}_i\rbrace,\tau) \approx  \sum_n e^{-i E_n \tau} \left[ \langle n |\psi_0 \rangle \right. \nonumber \\
&+& 2 \left. i \exp\left(i \frac{v}{2} \frac{\lambda_0}{4 \pi a} \log(\pi/2-\tau)\right)  \right. \nonumber \\
&  \cdot& \left. \sin\left( \frac{v}{2}\frac{\lambda_0}{4 \pi a} \log(\pi/2-\tau) \right)  \langle n | v \rangle \langle v | \psi_0 \rangle \phi_n( \lbrace \vec{x}_i \rbrace ) \right]\nonumber \\
\label{eq:final_c_r}
\end{eqnarray}

\begin{figure}
\includegraphics[scale=0.35]{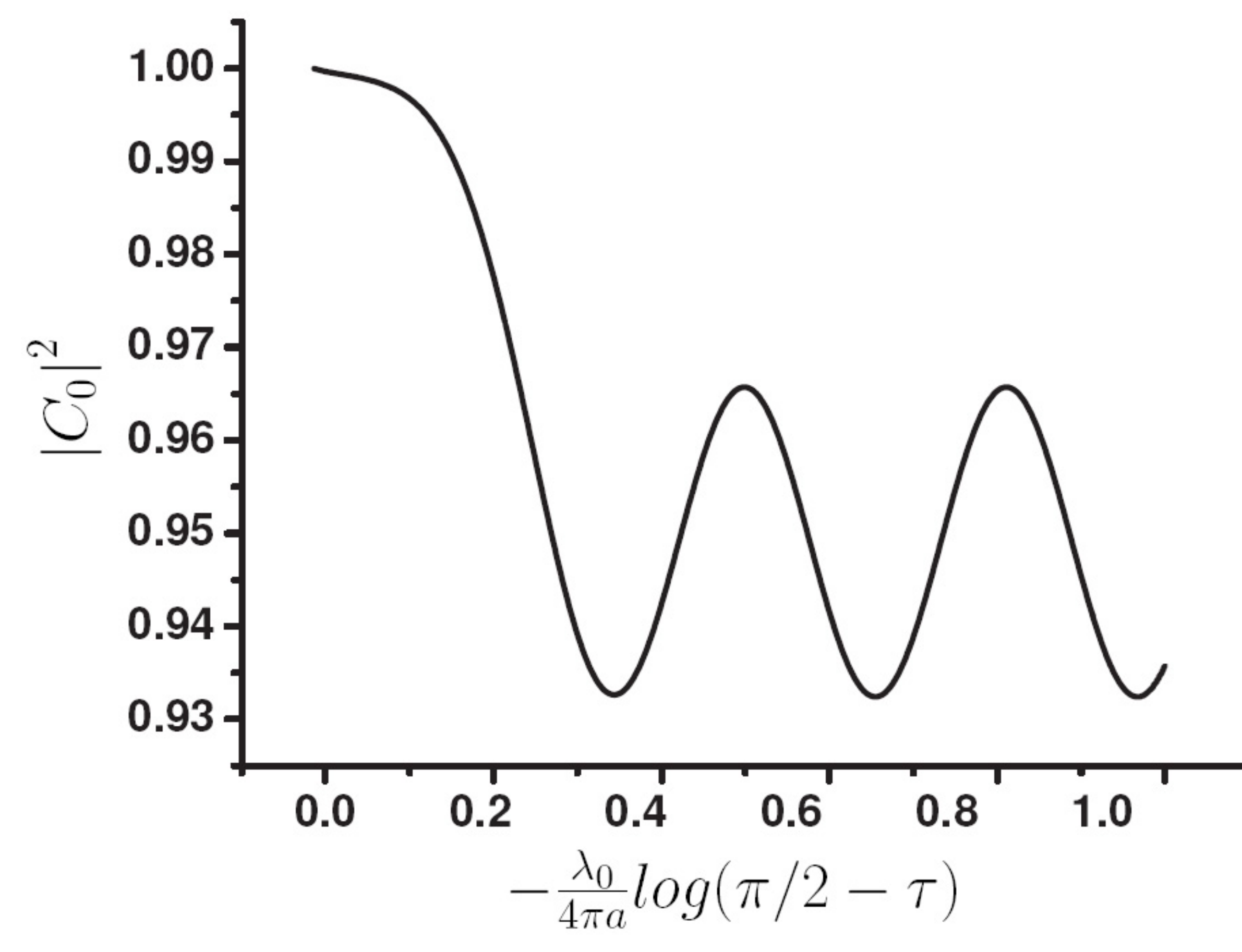}
\caption{The probability for the particle to remain in the resonant ground state in the co-moving frame, for large finite scattering lengths, $a \gg \lambda_0$, as a function of $-\lambda_0/(4 \pi a) \log(\pi/2-\tau)$. This result has been obtained by numerically solving Eq.~(\ref{eq:Int_Picture_definitions}) with $\lambda_0/(4\pi a) = 0.015$ and $r_0/\lambda_0=10^{-3/2}$. The system is initially prepared in the ground state of the resonant model. Very quickly the probability satisfies Eq.~(\ref{eq:final_c_r}), and develops oscillations at the frequency $v/2 = 20.14$.}
\label{fig:prob}
\end{figure}

\noindent The first term in Eq.~(\ref{eq:final_c_r}) is the initial expansion coefficients, and it produces the scale invariant dynamics. The presence of the second term will produce a beat in the probability density of the form $\sin^2(v \lambda_0/(4\pi a) \log(\pi/2-\tau)/2)$, in the co-moving frame, or $\sin^2(v \lambda_0/(4 \pi a) \log(t/\lambda_0^2)/2)$, in the laboratory frame. The beat is illustrated in Fig.~(\ref{fig:prob}), where we show the numerical solution for the probability to be in the resonant ground state, $|C_0|^2$, for a wave function initially in the resonant ground state. The solution is obtained by numerically solving the time dependent Schrodinger equation, Eq.~(\ref{eq:nirf_se}). The oscillations at frequency, $v/2$, are easily observed and are well described by Eq.~(\ref{eq:final_c_r}).

As an example of the breaking of scale invariance, consider the time evolution of the moment of inertia for the two-body system:

\begin{equation}
\lim_{t \rightarrow \infty} \langle r^2 \rangle (t) \approx \lambda^2(t)\left( \langle X^2 \rangle(\pi/2) + \langle x^2 \rangle(\tau(t) \right).
\end{equation}

\begin{figure}
\includegraphics[scale=0.25]{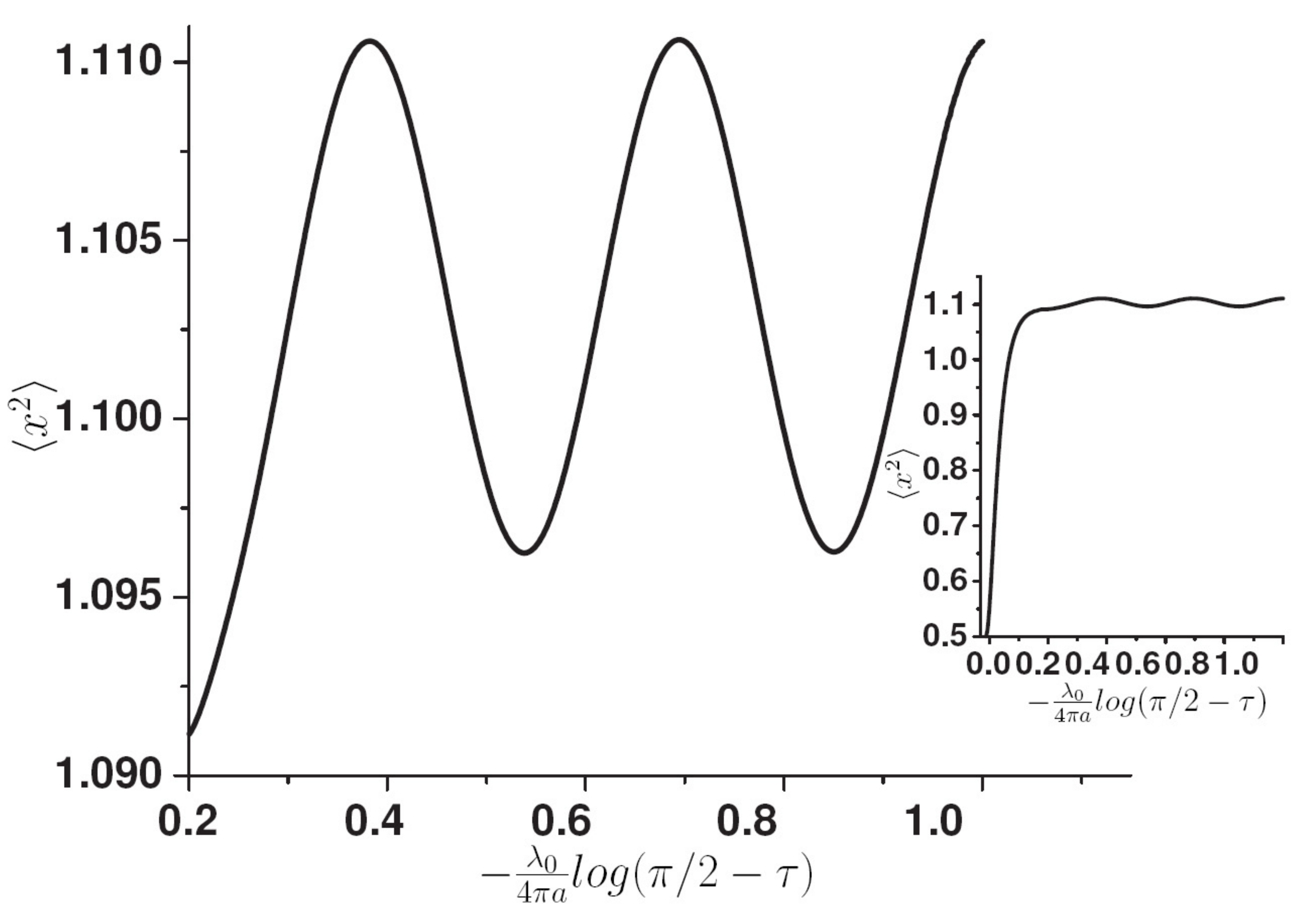}
\caption{The time evolution of $\langle x^2 \rangle(\tau)$ as a function of $-\lambda_0/(4\pi a) \log(\pi/2-\tau)$. This has been obtained by numerically solving the near resonant wave function, Eq.~(\ref{eq:Int_Picture_definitions}). In this calculation, the system was prepared in the ground state with $\lambda_0/(4\pi a) = 0.015$ and $r_0/\lambda_0 = 10^{-3/2}$.   The dynamics can be fit to Eq.~(\ref{eq:xs_dynamics}), with oscillations at frequency $v/2= 20.14$. In the inset, the dynamics over the entire range is shown.} 
\label{fig:rsquared}
\end{figure}

\noindent In the co-moving frame, the contribution to the center of mass, $\langle X^2 \rangle(\tau)$, will saturate for the product states under consideration here. In this case, only the relative motion is complicated due to the breaking of scale invariance. In Fig.~(\ref{fig:rsquared}), we show the relative moment of inertia. In the co-moving frame, the log-periodic oscillations are clearly visible. In terms of the laboratory frame, the dynamics of the relative moment of inertia are given by:

\begin{align}
\lim_{t \rightarrow \infty}\langle x^2 \rangle(t)  &\approx A + B \sin\left(v \frac{\lambda_0}{4 \pi a} \log(t/\lambda_0^2) \right) \frac{\lambda_0^2}{t}  \nonumber \\
&+ D \sin^2\left(  \frac{v}{2} \ \frac{\lambda_0}{4 \pi a} \log(t/\lambda_0^2) \right).
\label{eq:xs_dynamics}
\end{align}

\noindent In Eq.~(\ref{eq:xs_dynamics}), the coefficients $A$, $B$, and $D$ depend on the initial conditions, and the range of the potential, $r_0$, through the state $| v \rangle$. Explicit expressions for these coefficients are given in Appendix \ref{app:coefficients}. This result is consistent with Eq.~(\ref{eq:physical_operator_scaling}) and is identical to the time dependent perturbation theory, Eq.~(\ref{eq:pt_xs}), if one expands the sine functions to first order in $v \lambda_0/(4\pi a) \log(t/\lambda_0^2)$. The presence of the beat in the expansion coefficients can lead to significant deviations from the scale invariant dynamics, which can be seen in Fig.~(\ref{fig:rsquared}).

In addition to these results, we note that the two-body interaction does not break translational invariance in the laboratory frame. Thus, it is meaningful to study the dynamics of the momentum distribution, $n(k,t)$. In particular, we examine the dynamics of the contact \cite{Tan08}. The contact is defined by the assymptotic behaviour of the momentum distribution: $C(t) = \lim_{k \rightarrow \infty} k^4 n(k,t)$. The dynamics of the contact at resonance, and for weak interactions has been discussed previously in Ref.~\cite{Qu16}. Here we extend their analysis to study the dynamics of the contact away from resonance.

In order to obtain the momentum distribution near resonance, we will again assume that the initial wave function is a product state between the center of mass, and relative motion. One can then integrate out the center of mass coordinate to unity, and examine the momentum distribution for the relative coordinate. The relative momentum distribution for this case is related to the Fourier transform of the solution for the relative wave function, Eq.~(\ref{eq:Int_Picture_definitions}), in the laboratory frame:

\begin{align}
\psi(k,t) = \sum_n & \frac{e^{-2 i n \tau}}{\lambda^{3/2}(t)} \langle n | e^{i \frac{\lambda_0}{a} \log(t/\lambda_0^2) \tilde{V}_I} |\psi_0\rangle  \cdot \nonumber \\
&\int d^3r e^{i \frac{r^2}{2} \frac{\dot{\lambda}(t)}{\lambda(t)} - i \vec{k} \cdot \vec{r}} \phi_n \left( \frac{\vec{r}}{\lambda(t)}\right),
\label{eq:fourier_transform}
\end{align}

\noindent For large momenta, the integrand of Eq.~(\ref{eq:fourier_transform}) will be dominated by the contribution at short distances, $r \ll k^{-1}$. Again it is possible to obtain an analytical expression for the contact:

\begin{align}
C(t) &= \lim_{k \rightarrow \infty} k^4 n(k,t) = \lim_{k \rightarrow \infty} k^4 \left| \psi(k,t) \right|^2,\nonumber \\
&= \frac{1}{\lambda(t)} \tilde{C}(t),\nonumber \\
\tilde{C}(t) &= \left|\sum_n \langle n | e^{i \frac{\lambda_0}{a} \log(t/\lambda_0^2) \tilde{V}_I} |\psi_0\rangle e^{-2 i n \tau} \frac{\sqrt{\pi}}{2} f_n \right|^2,
\label{eq:contact_formula}
\end{align}

\noindent where the matrix elements of $\tilde{V}_I$ and $f_n$ are given in Eq.~(\ref{eq:V_tilde}).

At resonance, the contact in the co-moving frame, $\tilde{C}(\tau)$, tends to a constant which depends on the initial conditions. This result was obtained in Ref.~\cite{Qu16}. Near resonance, however, the expansion coefficients are time dependent due to the log-periodic beat, Eq.~(\ref{eq:final_c_r}). The beat in  the expansion coefficients will translate to a beat in the contact:

\begin{align}
\lim_{t\rightarrow \infty}C(t) \approx \frac{E}{\lambda(t)} + \frac{F}{\lambda(t)} \sin^2\left(\frac{v}{2} \ \frac{\lambda_0}{4 \pi a} \log\left(\frac{t}{\lambda_0^2}\right) \right), \nonumber \\
\label{eq:contact_dynamics}
\end{align} 

\noindent where $E$ and $F$ are two coefficients, which are given explicitly in Appendix \ref{app:coefficients}, and we have neglected terms that vanish as $\lambda_0^2/t$. The first term is the resonant scale invariant result, while the second term is the deviation that arises from breaking the scale invariance. Again, the presence of the log-periodic beat only depends on the deviation from resonance. The amplitude of the beat is controlled by the constant $F$ which depends on the finite range, $r_0$, through the state $|v\rangle$.

\subsection{Weakly Interacting}

At small scattering lengths, we expand the total Hamiltonian around the non-interacting fixed point. As seen in Sec.~\ref{sec:dynamics_near_formalism}, the scaling of the interaction is: $\alpha = -1$. In this case the dynamics in the co-moving frame  are well described by time dependent perturbation theory. In the laboratory frame the dynamics will become a simple time-dependent rescaling in the long time limit.

\subsection{Experimental Application}

\begin{figure}
\includegraphics[scale=0.40,trim = {0 1cm 0 0},clip]{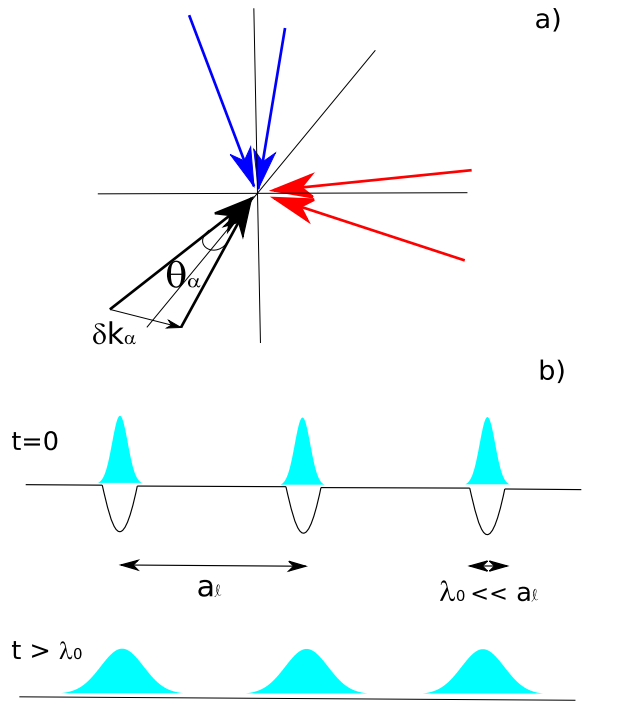}
\caption{Proposed experimental set up for examining broken scale invariance on two-body dynamics. a) To create a three dimensional lattice, with a lattice constant, $a_{l}$, much larger than the optical wavelength, three pairs of coplanar beams are needed. Each pair of beams will have the same wave number, $k$, but an angle, $\theta$ between them. The resulting optical lattice will be due to the difference between the two beams: $\delta k_{\alpha} = \sin(\theta_{\alpha}/2) k_{\alpha}$, for $\alpha = x,y,z$. If each of the different pairs of beams lie in different orthogonal planes, the result will be a square lattice. b) A schematic of the experiment. Here we show a single dimension of the resulting optical lattice. At $t=0$ the lattice is removed so the two-body systems can expand in free space. For times,  $\lambda_0^2 \ll t \ll a_{l} \lambda_0$, the dynamics of the whole system will be equivalent to an ensemble of independent two-body systems.}
\label{fig:experiment}
\end{figure}

Although the conclusion of the relevance of the perturbation near the resonant fixed point is general, and applicable to both few and many body systems, here we discuss a specific experiment to verify our theoretical predictions on broken scale invariant dynamics. We propose to prepare an ensemble of two-body states tightly confined in micro-traps that are periodically arranged in an optical lattice (see Fig.~(\ref{fig:experiment})). The confinement radius, $\lambda_0$, of each micro-trap, to a very good approximation, is simply the harmonic length of each tightly confining micro-trap, which depends on the laser intensities: $\lambda_0 =\sqrt{\hbar/2m \omega}$, and $\omega^2=(1/3m)\nabla^2 V({\bf r})|_{\vec{r}=\vec{r}_0}$ (assuming cubic lattice symmetry), where the derivatives are evaluated at the lattice sites of the optical lattice, which are formed by the minima of the confining potential energy, $V(\vec{r})$, see Eq.~(\ref{V}). To enhance the effect of broken scale invariance and for the convenience of experimental observation, we further propose to use three pairs of coplanar lasers angled at a small $\theta$ to create a lattice with a controllable and relatively large lattice constant, $a_l(\theta)$. 

The laser set up is shown in Fig.~(\ref{fig:experiment}) a). For each dimension, the two coplanar beams have wave vectors  $\vec{k}_1$, and $\vec{k}_2$. The two beams will then interfere and create a standing wave with an effective wave vector, $\delta \vec{k}_{\alpha} = (\vec{k}_1 - \vec{k}_2)_{\alpha} = k \sin(\theta/2) \hat{e}_{\alpha}$, where $\hat{e}_{\alpha}$ is the unit vector for direction $\alpha = x,y,z$. In this experiment we assume that each pair of coplanar beams are constructed to produce a periodic potential of the form:

\begin{align}
V(\vec{r}) &= \nonumber \\
&V_0 \cos^2\left(k \sin\frac{\theta}{2} x \right)\cos^2\left(k \sin\frac{\theta}{2} y \right)\cos^2\left(k \sin\frac{\theta}{2} z \right)
\label{V}
\end{align}

\noindent where $V_0$ is proportional to the laser intensity. The lattice constant for this potential is given by:

\begin{equation}
a_l = \frac{\pi}{k \sin\frac{\theta}{2}}.
\end{equation}

\noindent In practice, the lattice constant can be tuned up to the order of millimetres by decreasing $\theta$.

Now it is possible to create an ensemble of two-body systems via an optical lattice \cite{Bloch05, few_body11} by downloading pre-cooled atoms either a) in the presence of Feshbach resonance, or b) in the absence of scattering. The two situations, a) and b), correspond to the two fixed points studied in previous sections, the resonant and free particle fixed points, respectively. As the tunnelling is negligible in the limit of tight confinement, each approximately harmonic micro-trap will host conformal tower states. If the atoms are at rest in the ground state, 
the initial state will be at the bottom of the tower.
At $t=0$, the lattice is then turned off but the magnetic field is simultaneously varied so that the systems are no longer at resonance (a), or zero scattering (b).

The expansion of the two-body ensemble, shown in Fig.~(\ref{fig:experiment}) b), can then be observed to verify the main conclusions about the relevance or irrelevance of deviations from a scale invariant fixed point. Near resonance the dynamics should be modified by a non-perturbative log-periodic time dependence, see Eqs.~(\ref{eq:final_c_r}) and (\ref{eq:xs_dynamics}) as well as Figs.~(\ref{fig:prob}) and (\ref{fig:rsquared}), while for weak interactions the long time dynamics are equivalent to a rescaling. As far as $\lambda(t) \ll a_{l}/2$, each two-body system will expand independently in free space. For systems with initially tight confinement, $\lambda_0 \ll a_{l}$, the effect of broken scale invariance on the  dynamics will then be visible for times: $\lambda_0^2 \ll t \ll a_{l} \lambda_0$.

\section{Dynamics of a Non-Interacting Quantum Gas in the Presence of an Impurity}
\label{sec:near_res_imp}

We now apply the formalism of Sec.~\ref{sec:dynamics_near_formalism} to the dynamics of a non-interacting quantum gas in the presence of a short ranged, immobile impurity, in three spatial dimensions. The effect of the impurity is to create an external potential for the quantum gas. The effective Hamiltonian for the system, is given by Eq.~(\ref{eq:nirf_se}), with the inter-particle interaction, $V(\vec{x})$, set to zero. 

As mentioned in the previous section, the relative motion of a two body system is equivalent to a single particle moving in an external potential. For that reason, the results from Sec.~\ref{sec:two_body} are equally applicable to a single particle interacting with an impurity potential, see Eqs.~(\ref{eq:final_c_r}), and (\ref{eq:xs_dynamics}) as well as Figs.~(\ref{fig:prob}) and (\ref{fig:rsquared}). Although these results were obtained for a single particle, the extension to $N$ particles is straightforward. For multiple cold atoms, the dynamics can be understood by examining how $N$ particles occupy the eigenstates of $\tilde{V}_I$. For fermions, the Pauli-exclusion principle prevents multiple fermions with the same spin to occupy the state, $| v\rangle$. This results in the appearance of a single frequency $v$ in the dynamics. On the other hand, if the system is composed of bosons which have condensed, the dynamics will be identical to a single particle, and will still have oscillations at the frequency, $v$.

In general, for a $N$-body system, the wave function in the co-moving frame can be written in terms of:

\begin{align}
\phi(x_1,...,x_N, & \tau) = \sum_{\lbrace n_i \rbrace} \left[ \psi({\lbrace n_i \rbrace},\tau)e^{-i \sum_{i=1}^N 2n_i \tau} \times \right. \nonumber \\
&\left.\frac{1}{\sqrt{N!}}\sum_P (\pm 1)^P \phi_{n_{P_1}}(x_1) \phi_{n_{P_2}}(x_2)...\phi_{n_{P_N}}(x_N) \right]
\end{align}

\noindent where the first summation runs over all distinct combinations  of single particle conformal states, $n_1,...,n_N$, where each individual $n_i$, runs from $n_i = 0,1,...,n_{max}$, and $i = 1,2,...,N$. The second summation is over all the permutations of the set $n_1,...,n_N$. The factor of $(\pm 1)^P$ ensures the correct particle exchange symmetry for either bosons, or fermions, with $P$ being the number of exchanges to reach the given permutation. The expansion coefficients, $\psi({\lbrace n_i \rbrace},\tau)$, are normalized to unity.

For this general wave function, one can show that the moment of inertia for a non-interacting quantum gas has the same form as Eq.~(\ref{eq:xs_dynamics}). In Appendix \ref{app:coefficients}, explicit expressions for the the coefficients, $A$, $B$, and $D$ are given for $N$ particles.

\subsection{Trapped Impurity}
\label{sec:mobile_impurity}

So far the impurity has been assumed to be immobile. Practically, an impurity atom can never be truly immobile thanks to zero point motion. Here we consider a more realistic set up: a quantum gas interacting with an impurity of mass, $M$, that is subjected to a harmonic trap of frequency $\omega_I$. 

In the laboratory frame, the Hamiltonian for the system is:

\begin{equation}
H = \sum_i -\frac{1}{2} \nabla_i^2 - \frac{1}{2 M}\nabla_I^2 + \frac{1}{2}M \omega_I R_I^2 + \sum_i U(\vec{r}_i - \vec{R}_I),
\label{eq:Mobile_H}
\end{equation}

\noindent where $U(\vec{r})$ is the short range atom-impurity interaction. In this Hamiltonian, the motion of the trapped impurity atom is included alongside the quantum gas. The operators, $R_I$ and $-i \nabla_I$ are the coordinate and momenta operators for the impurity atom, while $\vec{r}_i$ and $-i \nabla_i$ are still the coordinates and momenta for the quantum gas. 

One can still use Eq.~(\ref{eq:Castin_wf}) to move to the expanding co-moving frame. In order to examine the dynamics of the quantum gas, it is ideal to choose Eq.~(\ref{eq:coordinates}) as the solution for $\lambda(t)$. This choice of $\lambda(t)$ then gives the modified effective Hamiltonian:

\begin{align}
\tilde{H} = &\sum_i \left( -\frac{1}{2} \tilde{\nabla}_i^2 + \frac{1}{2} x_i^2 + \lambda^2(\tau) U(\lambda(\tau)(\vec{x}_i - \vec{X}_I))\right) \nonumber \\
-& \frac{1}{2 M}\tilde{\nabla}_I^2 + \frac{1}{2}M X_I^2\left( \omega_I^2 \lambda^4(\tau) + 1\right)
\label{eq:mobile_imp_H}
\end{align}

\noindent where $\vec{X}_I = \vec{R}_I/\lambda(t)$.

Eq.~(\ref{eq:mobile_imp_H}) states that the impurity is subject to a harmonic trap of frequency: $\sqrt{\omega_I^2 \lambda^4(\tau) + 1}$. As $\tau$ approaches $\pi/2$, the frequency of the trap will diverge. In this case, we expect that the motion of the impurity to be adiabatic, and that the impurity will become more and more localized near the origin.

\begin{figure}
\includegraphics[scale=0.4]{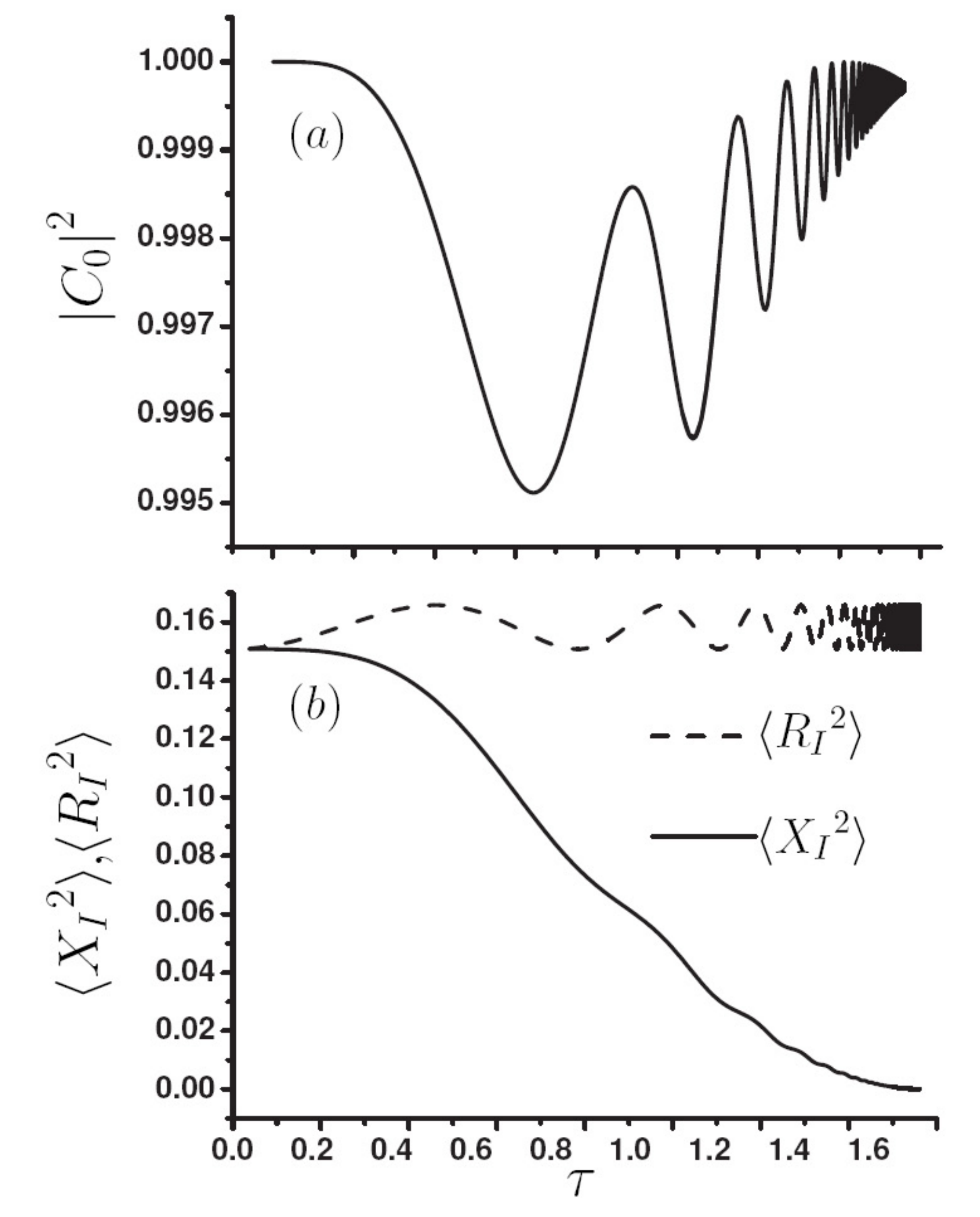}
\caption{The time evolution of the trapped impurity according to Eq.~(\ref{eq:mobile_imp_H}) in units of $\lambda_0$, for $M \omega_I \lambda_0^2 = 3$. a) The probability of being in the instantaneous ground state of Eq.~(\ref{eq:mobile_imp_H}). b) The fluctuations of the trapped impurity position, Eq.~(\ref{eq:Impurity_Position}), in both the laboratory (dashed line) and co-moving frames (solid line).}
\label{fig:impurity}
\end{figure}

To test this hypothesis, we show in Fig.~(\ref{fig:impurity}) a) the probability for the impurity to be in the instantaneous ground state of the trap, when it was initially prepared in the ground state. It is easy to see that the adiabatic approximation works extremely well in the long time limit.  The fluctuations in the position of the trapped impurity, $\sqrt{\langle X_I^2 \rangle}$, can then be related to the instantaneous trap size:

\begin{equation}
\sqrt{\langle X_I^2 \rangle} = \sqrt{\frac{1}{M \omega_I \lambda_0^2}} \cos(\tau),
\label{eq:Impurity_Position}
\end{equation}

\noindent which vanishes in the long time limit. The dynamics of this expectation value is shown in Fig.~(\ref{fig:impurity}) b).

The adiabatic dynamics in the co-moving frame is intuitive when one considers the motion in the laboratory frame. For heavy impurities, or large trapping potentials, the interaction between the impurity and the gas will not excite the impurity. As a result,  when the gas expands further away from the impurity, the impurity will remain near the origin, and it's exact position in the trap will become more and more irrelevant to the expanding gas. 

The adiabaticity of the impurity results in a form of coarse grained dynamics for the quantum gas. In the long time limit, the dynamics of the quantum gas will be insensitive to the initial preparation of the impurity, and all the results obtained for the immobile impurity will be valid for the trapped impurity when: $t \gg \lambda_0^2$.

\section{Discussion}
\label{sec:disc}

The main result of this work is the formalism developed to describe the dynamics slightly away from different scale invariant fixed points, see Eqs.~(\ref{eq:deviation_scaling}), (\ref{eq:Int_Picture_definitions}), (\ref{eq:T_form}), (\ref{eq:V_matrix}), and (\ref{eq:physical_operator_scaling}). In the context of cold gases, this approach is equally valid near both the non- and resonantly-interacting fixed points. The dynamics near these fixed points can be qualitatively determined by the scaling of the deviation, $\alpha$, given in  Eq.~(\ref{eq:deviation_scaling}). This scaling can be determined by using a scaling analysis, and is found to be given by the linearized beta function near a given fixed point. For deviations that have a scaling, $\alpha \geq 1$, the long time dynamics, $t \gg \lambda_0^2$, can not be described by time dependent perturbation theory. Instead, the long time  behaviour of the wave function is encapsulated in a time-independent, universal matrix, $\tilde{V}_I$, given in Eq.~(\ref{eq:V_matrix}). In the opposite limit, $\alpha < 1$, perturbation theory provides an accurate description of the dynamics. In this case, the wave function will freeze in the co-moving reference frame, while the dynamics in the laboratory frame will become a simple time-dependent rescaling.

This formalism was applied to the dynamics of both the non-interacting quantum gas in the presence of an impurity, and of the interacting two-body problem in three spatial dimensions. Near resonance, a log-periodic beat appears in the expansion coefficients, Eq.~(\ref{eq:final_c_r}), with a frequency which depends on the finite range of the interaction, Eq.~(\ref{eq:v}). This beat will is visible in the dynamics of the system, see Eqs.~(\ref{eq:xs_dynamics}), and  (\ref{eq:contact_dynamics}).

The relevancy of the interaction also allows one to compare the results discussed above with one dimensional systems with either s-wave, or p-wave, interactions. For one dimensional s-wave systems, the non-interacting fixed point and resonant fixed point switch; i.e. the resonant fixed point is infra-red stable, while the non-interacting fixed point is unstable. Repeating our calculation shows the dynamics agree with the relevancy of the perturbation. Namely, the scaling dimension of the deviation near the weakly interacting fixed point have scaling, $\alpha = 1$, and are thus relevant, while $\alpha = -1$, near the resonant fixed point. For p-wave interactions in one dimension, a renomalization group approach shows the physics should be identical to the three dimensional s-wave. We have repeated our calculation for p-wave interactions in one dimension and found that the dynamics are identical to the three dimensional s-wave. All the results obtained for three dimensions will then apply to one dimensional systems.

The method employed in this work is general and can be applied to a wide variety of scale invariant and nearly scale invariant quantum systems. However, with increasing system size, the matrix $\tilde{V}_I$, given in Eq.~(\ref{eq:V_matrix}), becomes exponentially more complex. Although this is a daunting task, our work shows that it is still possible to make scaling arguments for the dynamics of strongly interacting quantum gases. We intend to examine the many body problem more thoroughly in a future work.

This work was funded by the National Science and Engineering Research Council of Canada (Contract No. 288179), and the Canadian Institute for Advanced Research. The authors would like to thank Shao-Jian Jiang for helpful discussions.

\appendix

\numberwithin{equation}{section}
\renewcommand\theequation{\Alph{section}.\arabic{equation}}

\section{Physics in the Non-Inertial Co-Moving Frame}
\label{app:comoving_frame}

In this appendix we discuss how to examine a given physical system in a non-inertial, expanding, co-moving frame. We begin by examining the time dependent many body Schrodinger equation for a system of $N$ particles with spin-independent interactions, $V(\vec{r})$, and in an external potential, $U(\vec{r})$, in three spatial dimensions:

\begin{align}
i &\partial_t \psi\left( \lbrace\vec{r}_i, \sigma_i \rbrace, t \right)= H  \psi\left( \lbrace\vec{r}_i, \sigma_i\rbrace, t\right), \nonumber \\
H &= \sum_i\left[ -\frac{1}{2} \nabla_i^2 + U(\vec{r}_i)  \right] +\frac{1}{2}\sum_{i,j} V(\vec{r}_i - \vec{r}_j),
\label{O:eq:Schrodinger} 
\end{align}

\noindent where the atomic mass, $m$, and $\hbar$ have been set to unity. $\vec{r}_i$ and $\sigma_i$ designate the position and spin of the $i$th particle, respectively.

For systems with scale invariance or that are nearly scale invariant, it is convenient to introduce the wave function:

\begin{eqnarray}
\psi(\lbrace\vec{r}_i, &&\sigma_i \rbrace, t) = \nonumber \\
&& \frac{1}{\lambda^{3N/2}(t)} e^{\frac{i}{2} \sum_ir_i^2 \dot{\lambda}(t)/\lambda(t)} \phi\left( \lbrace \frac{\vec{r}_i}{\lambda(t)},\sigma_i \rbrace, \tau(t) \right), \nonumber \\
\label{O:eq:Castin_wf}
\end{eqnarray}

\noindent where $\lambda(t)$ is a time dependent scaling factor with units of length, and $\tau(t)$ is a dimensionless effective time. The wave function, $\phi(\lbrace \vec{r}_i / \lambda(t),\sigma_i \rbrace, \tau(t))$, contains all the many body information of the original wave function, and is both appropriately symmetrized and normalized.

After substituting Eq.~(\ref{O:eq:Castin_wf}) into the time dependent Schrodinger equation, Eq.~(\ref{O:eq:Schrodinger}), one finds a Schrodinger-like equation for the field $\phi(\lbrace\vec{r}_i / \lambda(t), \sigma_i\rbrace,\tau)$:

\begin{widetext}
\begin{equation}
i \frac{\partial \tau}{\partial t} \frac{\partial}{\partial \tau} \phi\left( \lbrace \vec{x}_i, \sigma_i \rbrace, \tau(t) \right) = \left( \sum_i \left[- \frac{1}{2} \frac{1}{\lambda^2(t)} \tilde{\nabla}_{i}^2 + \frac{x_i^2}{2} \ddot{\lambda}(t)\lambda(t) + U(\lambda(t) \ \vec{x}_i) \right] + \frac{1}{2}\sum_{i,j} V(\lambda(t) \ (\vec{x}_i-\vec{x}_j)) \right)\phi\left( \lbrace \vec{x}_i, \sigma_i \rbrace, \tau(t) \right).
\label{O:eq:phi_se}
\end{equation}
\end{widetext}

\noindent The effective spatial and temporal coordinates in Eq.~(\ref{O:eq:phi_se}) are defined as:

\begin{align}
\vec{x}_i &= \vec{r}_i /\lambda(t), & i &=1,2,...,N & \tau(t),
\label{O:eq:coordinates}
\end{align}

\noindent respectively, and the operator, $\tilde{\nabla}_i$, acts on the effective position, $\vec{x}_i$. In order for the dynamics after the transformation, Eq.~(\ref{O:eq:phi_se}), to be identical to a Schrodinger equation, the time derivative and the kinetic energy must scale in the same way. For this reason, it is necessary to choose:

\begin{equation}
\frac{\partial \tau(t)}{\partial t} = \frac{1}{\lambda^2(t)}.
\label{O:eq:def_tau_t}
\end{equation}

\noindent One can then derive a Schrodinger equation for the field $\phi(\lbrace \vec{x}_i, \sigma_i \rbrace, \tau)$ in terms of the effective coordinates:

\begin{align}
& i \frac{\partial}{\partial \tau} \phi(\lbrace \vec{x}_i, \sigma_i \rbrace, \tau) = \tilde{H}(\tau) \phi(\lbrace \vec{x}_i, \sigma_i \rbrace, \tau), \nonumber \\
& \tilde{H}(\tau) = \sum_i\left[-\frac{1}{2}\tilde{\nabla_i}^2 + \frac{x_i^2}{2} \ddot{\lambda}(t(\tau)) \lambda^3(t(\tau)) + \lambda^2(\tau) U(\lambda(\tau) \vec{x}_i)\right] \nonumber \\
&+ \frac{1}{2}\sum_{i,j}\lambda^2(\tau) V(\lambda(\tau)(\vec{x}_i - \vec{x}_j)).
\label{O:eq:intermediate_se}
\end{align}

At this stage, the choice of $\lambda(t)$ is arbitrary.  The optimal choice of $\lambda(t)$ for the present discussion is to eliminate the time dependence in the harmonic term by setting: 

\begin{align}
\ddot{\lambda}(t)\lambda^3(t) &=1, & \lambda(0) &= \lambda_0, & \dot{\lambda}(0) &= 0,
\label{O:eq:differntial_eqn_lambda}
\end{align}

\noindent  where $\lambda_0$ is the length scale describing the initial wave function. The initial conditions are chosen such that the initial wave function in the laboratory frame is related to the initial effective wave function via: $\psi(\lbrace \vec{r}_i\rbrace, 0) = \lambda_0^{-3N/2}\phi(\lbrace\vec{r}_i / \lambda_0\rbrace,0)$. The solution to Eqs.~(\ref{O:eq:def_tau_t}) and  (\ref{O:eq:differntial_eqn_lambda}) is:

\begin{align}
\lambda(t) &= \lambda_0 \sqrt{1 + \frac{t^2}{\lambda_0^4}},  & \tau(t) &= \arctan\left(\frac{t}{\lambda_0^2}\right), \nonumber \\
\lambda(\tau) &= \lambda_0 \sec(\tau), &  0&\leq  \tau < \pi/2.
\label{O:eq:lambda_tau}
\end{align}

\noindent Substituting the solution for $\lambda(t)$ into Eq.~(\ref{O:eq:intermediate_se}) gives the effective Schrodinger equation:

\begin{align}
& i \frac{\partial}{\partial \tau} \phi(\lbrace \vec{x}_i, \sigma_i \rbrace, \tau) = \tilde{H}(\tau) \phi(\lbrace \vec{x}_i, \sigma_i \rbrace, \tau), \nonumber \\
& \tilde{H}(\tau) = \sum_i\left[-\frac{1}{2}\tilde{\nabla}_i^2 + \frac{x_i^2}{2} + \lambda^2(\tau) U(\lambda(\tau) \vec{x}_i)\right] \nonumber \\
&+ \frac{1}{2}\sum_{i,j}\lambda^2(\tau) V(\lambda(\tau)(\vec{x}_i - \vec{x}_j)).
\label{O:eq:nirf_se}
\end{align}

\noindent This is just the Schrodinger equation for an interacting quantum gas in a harmonic trap with time dependent interactions.

\section{The Scaling of the Deviation from a Given Fixed Point}
\label{app:alpha}

In this appendix we show how to relate the scaling of the deviation from a given fixed point, $\alpha$, to the beta function linearized near the given fixed point. As $\alpha$ is critical in determining the relevancy of the deviation, relating it to the beta function will allow for a scaling analysis of non-equilibrium physics.

Consider a $d$-dimensional gas of $N$ atoms interacting via a short ranged, s-wave interaction:

\begin{equation}
V= g(\Lambda) \int_{\Lambda^{-1}} d^dr \psi^{\dagger}(\vec{r})\psi^{\dagger}(\vec{r})\psi(\vec{r})\psi(\vec{r})
\end{equation}

\noindent where $\psi(\vec{r})$ is the fermionic field operator, and $\Lambda$ is the UV cut-off for the theory.

Under a rescaling of the UV cut-off, $\Lambda' = \Lambda e^b$, where $b$ is the scaling factor. The  strength of the interaction will change. The change is given by the beta function:

\begin{align}
\beta(\tilde{g}(\Lambda)) = (d-2)\tilde{g}(\Lambda) + \tilde{g}^2(\Lambda),
\end{align}

\noindent where $\tilde{g}(\Lambda) = C_d g(\Lambda) \Lambda^{d-2}$, is the rescaled interaction strength \cite{Wilson83,Sachdev}, and $C_d= 2 \pi^{d/2} / (2\pi)^d\Gamma(d/2)$ is a constant that depends on the dimension. This is equivalent to Eq.~(\ref{eq:RG}) in the main text,  For this theory, the beta function has two zeros:

\begin{equation}
\tilde{g}^*(\Lambda)  =
\begin{cases}
0, \\
(2-d).
\end{cases}
\end{equation}

\noindent These two fixed points correspond to the non-interacting and resonantly-interacting theories, respectively for $d>2$.

At this point, one can easily write the deviation from scale invariance as:

\begin{align}
\delta H &= \left(g(\Lambda) - g^*(\Lambda) \right) h(\Lambda)
\end{align}

\noindent where $h(\Lambda)$ is the operator:

\begin{equation}
h(\Lambda) = \int_{\Lambda^{-1}} d^dr \psi^{\dagger}(\vec{r})\psi^{\dagger}(\vec{r})\psi(\vec{r})\psi(\vec{r}).
\end{equation}

In order to derive the scaling of the deviation, it is convenient to introduce a physical length scale, $a$:

\begin{equation}
a = \Lambda^{-1} f(\tilde{g}(\Lambda)).
\end{equation}

\noindent Since the theory is renormalizable, this length scale will not depend on the UV cut-off:

\begin{align}
\frac{\partial a}{\partial \log(\Lambda)} &= -a + \frac{\partial a}{\partial \tilde{g}(\Lambda)} \beta(\tilde{g}(\Lambda))=0. \nonumber \\
a &= \frac{\partial a}{\partial \tilde{g}(\Lambda)}\beta(\tilde{g}(\Lambda))
\end{align}

\noindent For small deviations from scale invariance, one can linearize the beta function near the fixed point, and integrate the above relation to obtain:

\begin{equation}
g(\Lambda) - g^*(\Lambda) \propto \frac{1}{\Lambda^{d-2}}\frac{1}{(\Lambda a)^{-\beta'(\tilde{g}^*(\Lambda))}}
\label{B:eq:dev}
\end{equation}

\noindent where

\begin{equation}
\beta'(\tilde{g}^*(\Lambda)) = \left.\frac{\partial \beta(\tilde{g}(\Lambda)}{\partial \tilde{g}(\Lambda)} \right|_{\tilde{g}(\Lambda) = \tilde{g}^*(\Lambda)}.
\end{equation}

Before transforming $\delta H$ to the co-moving frame, it is helpful to see how $\delta H$ rescales under the rescaling  $\psi(r)\rightarrow \lambda(t)^{-d/2} \psi(r/\lambda(t))$. In this case, by changing the variables: $x = r/\lambda(\tau)$, and $\tilde{\Lambda} = \lambda(\tau) \Lambda$, one can show that the interaction, in the laboratory frame, changes as:

\begin{align}
\delta H(\Lambda) \rightarrow \delta H'(\Lambda)  &=  \left(\frac{1}{\lambda(\tau)} \right)^{2- \alpha} \frac{1}{a^{\alpha}} \tilde{\Lambda}^{2-d-\alpha} h(\tilde{\Lambda}) \nonumber \\
&= \left(\frac{1}{\lambda^(\tau)}\right)^{2-\alpha} \delta H(\tilde{\Lambda}),
\end{align}

\noindent where we have defined:

\begin{equation}
\alpha = - \beta'(\tilde{g}^*(\Lambda)).
\end{equation}  

In moving to the co-moving frame, it is necessary to measure time in units of $\tau$. This  is done by multiplying this rescaled deviation by $\lambda^2(\tau)$, see Sec. \ref{sec:dynamics_near_formalism}. The result is that the deviation in the co-moving frame will have the form:

\begin{align}
\delta \tilde{H}(\tau) &= \lambda^2(\tau) \delta H'(\Lambda) \nonumber \\
&= \left( \frac{\lambda(\tau)}{a} \right)^{\alpha} \tilde{h}.
\end{align}

\noindent Here we note that, $ \tilde{h} = \tilde{\Lambda}^{2-d-\alpha} h(\tilde{\Lambda})$,  is a regularized matrix in the co-moving frame, which is true for the cases discussed in this article.

This expression is general and only depends on the beta function, and hence, the universality class of the fixed point. In terms of the s-wave scattering in three spatial dimensions, one can show that near resonance: $\alpha = 1$, while for the non-interacting case: $\alpha = -1$. These two scalings are identical to the scaling analysis presented in Sec.~\ref{sec:SI}. In Appendix \ref{app:derivation}, explicit calculations for the s-wave scattering for the two-body system, and the one body system in the presence of an impurity potential are shown. These explicit calculations are shown to be consistent with the scaling analysis presented here.

\section{Derivation of the Deviation Operator}
\label{app:derivation}

In this appendix we derive the deviation, $\delta \tilde{H}(\tau)$, from the transformed scale invariant Hamiltonian, $\tilde{H}$, in the expanding co-moving frame. The approach we employ here is applicable for both the non-interacting quantum gas with an impurity, and the relative dynamics of the two-body problem. The only difference is that in the two-body problem, one uses the reduced mass for the two particles. In both cases, the physical interaction will be some short ranged, spherically symmetric potential, $V(r)$. 

We begin with the radial Schrodinger equation in the co-moving frame:

\begin{align}
&i \partial_{\tau} \chi_l(x, \tau) = \tilde{H} \chi_l(x, \tau), \nonumber \\
& \tilde{H} = -\frac{1}{2}\partial_x^2 + \frac{1}{2}x^2 + \frac{l(l+1)}{2x^2}+ \lambda^2(\tau)V(x \lambda(\tau)), \nonumber \\
&\lambda(\tau) = \lambda_0 \sec(\tau),
\label{A:eq:nirf_se}
\end{align}

\noindent where we have set the (reduced) mass to unity, and $Y_{l,m}(\vec{x})$ is the spherical harmonic with angular quantum number $l$ and projection quantum number, $m$. The radial wave function, $\chi_l(x,\tau)$, is related to the full wave function via: $\phi_{l,m}(\vec{x},\tau) = Y_{l,m}(\hat{x}) \chi_l(x,\tau)/x$, and is properly normalized:

\begin{equation}
\int_0^{\infty}dx |\chi_l(x,\tau)|^2 =1.
\end{equation}

\noindent In what follows we will only  focus on the zero angular momentum, or s-wave, scattering of this potential, as higher angular momentum scattering is suppressed by a factor of $(\sqrt{E} r_0)^{2l}$, where $r_0$ is the range of the potential, and $E$ is the energy.

For specificity, we will consider the potential to be a square well of depth: $V_0 \lambda^2(\tau)$, and range: $r_0 / \lambda(\tau)$. This potential is consistent with the time dependence of $\lambda^2(\tau)V(x \lambda(\tau))$, and captures all the essential physics at low energies. It is important to note that the range and depth of the potential are changing at a rate set by $\lambda(\tau)$, which is much slower than the energy scale set by the finite range of the potential, $r_0$. This implies that we can use the adiabatic approximation. In this approximation, the effect of the finite scattering length is to impose the time dependent boundary condition at the range of the potential \cite{Tan08}:

\begin{equation}
\frac{\chi'(r_0/\lambda(\tau))}{\chi(r_0/\lambda(\tau))} = -\frac{\lambda(\tau)}{a}.
\label{A:eq:bethe-peierls}
\end{equation}



As discussed in the main text, we split the effective Hamiltonian in the co-moving frame, $\tilde{H}(\tau)$, into the effective Hamiltonian at the scale invariant fixed point, $\tilde{H}$, and a deviation, $\delta \tilde{H}(\tau)$:

\begin{align}
\tilde{H}(\tau) &= \tilde{H} + \delta \tilde{H}(\tau), \nonumber \\
\tilde{H} &= -\frac{1}{2} \partial_x^2 + \frac{1}{2}x^2 + \lambda^2(\tau) V_s (\lambda(\tau) x),
\end{align}

\noindent and $V_s(x)$ is a scale invariant potential. In this analysis the quantities of interest are the matrix elements of the deviation:

\begin{align}
\delta \tilde{H}(\tau) &=\lambda^2(\tau) V(x \lambda(\tau)) - \lambda^2(\tau)V_{s}( \lambda(\tau) x), \nonumber \\ 
\delta \tilde{H}(\tau) &=\lambda^2(\tau) V(x \lambda(\tau)) - V_{s}(x),
\label{A:eq:deviation_operator}
\end{align}

\noindent with respect to the eigenstates of $\tilde{H}$. In Eq.~(\ref{A:eq:deviation_operator}) we have used the fact that the system possesses scale invariance at a fixed point, i.e. $V_{s}(\lambda x) = \lambda^{-2} V_{s}(x)$. For more discussions on the role of scale invariance on the dynamics in the co-moving frame, see Sec.~\ref{sec:SI}.

We first evaluate the deviation from the resonant fixed point. The matrix elements of Eq.~(\ref{A:eq:deviation_operator}) near resonance can be determined by examining the zero angular momentum Schrodinger equation at, and near, resonance:

\begin{align}
E_{r,n} \chi_{r,n}(x) &= \left(-\frac{1}{2} \partial_x^2 +\frac{1}{2}x^2 + V_{res}(x) \right) \chi_{r,n}(x), \nonumber \\
E_{m}(\tau) \chi_m(x,\tau) &= \left(-\frac{1}{2} \partial_x^2 +\frac{1}{2}x^2 + \lambda^2(\tau)V\left(\frac{x}{\lambda(\tau)}\right) \right) \chi_{m}(x,\tau).
\label{A:eq:schrodinger_eqns}
\end{align}

\noindent The top and bottom lines correspond to the resonant and off resonant Schrodinger equations, respectively. The states $\chi_{r,n}(x)$ and $\chi_m(x,\tau)$ are the eigenstates of the system with energy $E_{r,n}$ and $E_m(\tau)$, and quantum numbers $n$ and $m$, for the resonant, and off resonant Hamiltonians, respectively.

At this stage one can multiply the resonant (off-resonant) Schrodinger equation by the state $\chi_m(x,\tau)$ ($\chi_{r,n}(x)$), and integrate over the range of the potential, $r_0 / \lambda(\tau)$.  The difference between the two Schrodinger equations is:

\begin{align}
&\int_0^{r_0/\lambda(\tau)} dx \chi_{r,n}(x) (E_m(\tau)-E_{r,n}) \chi_m(x,\tau) = \nonumber \\
&\int_0^{r_0/\lambda(\tau)} dx  \chi_{r,n}(x) \left[-\frac{1}{2} \partial_x^2 + \frac{1}{2}x^2 + \lambda^2(\tau) V(x \lambda(\tau)) \right] \chi_m(x,\tau) \nonumber \\
&- \int_0^{r_0/\lambda(\tau)} dx  \chi_{m}(x,\tau) \left[-\frac{1}{2} \partial_x^2 + \frac{1}{2}x^2 + V_{res}(x) \right] \chi_{r,n}(x)
\label{A:eq:difference_of_ses}
\end{align}

\noindent To obtain the deviation operator, we expand the difference to first order in $1/a$. To this order the expansion of Eq.~(\ref{A:eq:difference_of_ses}) gives:

\begin{align}
&\int_0^{r_0/\lambda(\tau)}dx (\lambda^2(\tau) V(x \lambda(\tau)) - V_{res}(x)) \chi_{r,m}(x) \chi_{r,n}(x) = \nonumber \\
&\int_0^{r_0/\lambda(\tau)} dx \left\lbrace (E_m(\tau) - E_{r,n}) \chi_{m}(x,\tau) \chi_{r,n}(x)\right\rbrace \nonumber \\
&-\frac{\lambda(\tau)}{2a} \chi_{r,m}(r_0/\lambda(\tau)) \chi_{r,n}(r_0/\lambda(\tau)).
\label{A:eq:difference_expansion}
\end{align}

\noindent In Eq.~(\ref{A:eq:difference_expansion}) we have used Eq.~(\ref{A:eq:bethe-peierls}) to evaluate the difference in kinetic energies.

In the second line of Eq.~(\ref{A:eq:difference_expansion}), the quantities $E_m(\tau)-E_{r,n}$ and $\chi_{r,n}(x) \chi_m(x,\tau)$  are to be expanded to first order in $\lambda(\tau)/a$. The integrals themselves will be proportional to $r_0 / \lambda(\tau)$, for $r_0 \ll \lambda_0$. Therefore the second line in Eq.~(\ref{A:eq:difference_expansion}) will be a correction of order $r_0 / a$ to the matrix elements, which is negligible in the large scattering length limit. After neglecting the terms proportional to $O(r_0/a)$, the expression for the deviation becomes:

\begin{align}
<m | &\delta\tilde{H}(\tau) |n>  = \nonumber \\
& \int_0^{r_0/\lambda(\tau)} \left( \lambda^2(\tau)V(x \lambda(\tau)) - V_{res}(x)\right) \chi_{r,m}(x) \chi_{r,n}(x) \nonumber \\
& = -\frac{\lambda(\tau)}{2 a} \chi_{r,m}(r_0/\lambda(\tau)) \chi_{r,n}(r_0/\lambda(\tau)).
\end{align}

To simplify the deviation further, we note that the resonant eigenstates outside the potential are given by:

\begin{align}
\chi_{r,n}(\vec{x})= \langle x |n\rangle &= \sqrt{2}\phi_{h.o,2n}(x), & E_{r,n} &= 2 n + 1/2,
\label{A:eq:r_eigenstates}
\end{align}

\noindent where $\phi_{h.o,n}(x)$ is the normalized one-dimensional harmonic oscillator wave function with quantum number $n= 0,1,2,...$, and with the harmonic length scale set to unity. For $x \ll 1$, the resonant eigenstates are constant near the origin. The continuity of the wave function at the boundary allows one to simplify the deviation to:

\begin{equation}
\langle m | \delta \tilde{H}(\tau) | n \rangle = -\frac{\lambda(\tau)}{2a} \chi_{r,m}(0)\chi_{r,n}(0).
\label{A:eq:matrix_elements_r}
\end{equation}

\noindent which is equivalent to the scaling analysis in Appendix \ref{app:alpha}.

For weak interactions, we compare the non- and weakly- interacting Schrodinger equations in the co-moving frame:

\begin{align}
E_{0,n} \chi_{0,n}(x) &= \left(-\frac{1}{2} \partial_x^2 +\frac{1}{2}x^2 \right) \chi_{0,n}(x), \nonumber \\
E_{m}(\tau) \chi_m(x,\tau) &= \left(-\frac{1}{2} \partial_x^2 +\frac{1}{2}x^2 \right. \nonumber \\
&\left. + \lambda^2(\tau)V(x  \lambda(\tau)) \right) \chi_{m}(x,\tau).
\end{align}

\noindent Here $\chi_{0,n}(x)$ is a non-interacting eigenstate with energy $E_{0,n}$ and quantum number $n = 0,1,2,...$:

\begin{align}
\chi_{0,n}(\vec{x}) &= \langle x |n \rangle = \sqrt{2} \phi_{h.o, 2n+1}(x), & E_{0,n} &=2n + 3/2. 
\end{align}

\noindent A calculation identical to the resonant case yields:

\begin{equation}
\langle m | \delta \tilde{H}(\tau)|  n \rangle = \frac{a}{2 \lambda(\tau)} \chi_{0,m}'(0) \chi_{0,n}'(0),
\label{A:eq:matrix_elements_non_int}
\end{equation}

\noindent which is again equivalent to the simple scaling analysis.

The approach used here is similar to Refs.~\cite{ Tan08,Zhang09}. The matrix elements of Eq.~(\ref{A:eq:matrix_elements_r}) and (\ref{A:eq:matrix_elements_non_int}) have an important connection to the thermodynamic contact first examined in Ref.~\cite{Tan08}. If one were to consider just the ground state expectation value of the deviation, Eq.~(\ref{A:eq:deviation_operator}), one would simply obtain the contact. Eq.~(\ref{A:eq:deviation_operator}) is then a natural extension of the idea of contact to a matrix. The relationship between the breaking of scale invariance and the contact in equilibrium physics has been discussed in Ref.~\cite{Hofmann12}.

\section{Solutions for the Expansion Coefficients near a Fixed Point}
\label{app:time_ordering}

In this section we analyse the time dependent expansion coefficients  near an arbitrary scale invariant fixed point. Near a fixed point, the effective wave function is given by:

\begin{eqnarray}
\phi(\lbrace \vec{x}_i \rbrace, \tau) &=& \sum_n C_n(\tau)e^{-i E_n \tau} \phi_{n}(\lbrace \vec{x}_i \rbrace), \nonumber  \\
C_n(\tau) &=& \langle n | T e^{-i \int_0^{\tau} d\tau' \delta \tilde{H}_I(\tau')} |\psi_0 \rangle, \nonumber \\
\delta \tilde{H}_I(\tau) &=& e^{i \tilde{H}_{s}\tau}\delta \tilde{H}(\tau) e^{-i \tilde{H}_{s} \tau}.
\label{B:eq:Int_Picture_definitions}
\end{eqnarray}

\noindent In Eq.~(\ref{B:eq:Int_Picture_definitions}), $| \psi_0 \rangle$ is the initial state, $\phi_n(\lbrace\vec{x}_i \rbrace) = \langle \lbrace x_i \rbrace | n \rangle$ is the eigenstate of the transformed scale invariant Hamiltonian, $\tilde{H}$, with energy, $E_n=2n$ and $n = 0,1,2,...$. The deviation from, $\tilde{H}$, is given by  $\delta \tilde{H}(\tau)$.

To evaluate the expansion coefficients, $C_n(\tau)$, it is necessary to expand the exponential in the second line of Eq.~(\ref{B:eq:Int_Picture_definitions}). Here we will examine the $mth$ order for the expansion coefficient:

\begin{align}
C^{(m)}_n(\tau) &= \frac{(-i)^m}{m!}\int_0^{\tau}d\tau_1 ...\int_0^{\tau}d\tau_m \nonumber \\
& \langle n| T \delta\tilde{H}_I(\tau_1)...\delta\tilde{H}_I(\tau_m) | \psi_0 \rangle.
\label{B:eq:coefficients_time_ordered}
\end{align}

We can remove the time ordering operator, $T$, by choosing a specific ordering of the time coordinates in Eq.~(\ref{B:eq:coefficients_time_ordered}). Each possible ordering of the coordinates will generate an equal contribution, eliminating the $m!$ factor in Eq.~(\ref{B:eq:coefficients_time_ordered}):

\begin{align}
C_n^{(m)}(\tau) &= -i^m \sum_{l_1,...l_{m}}\int_0^{\tau}d\tau_1 ...\int_0^{\tau_{m-1}}d\tau_m \nonumber \\
& \langle n| \delta \tilde{H}(\tau_1) | l_1 \rangle ... \langle l_{m-1} | \delta \tilde{H}(\tau_m) | l_m \rangle \cdot \nonumber \\
& e^{2i(n-l_1)} ... e^{2i(l_{m-1}-l_m)} C_{l_m}(0),
\label{B:eq:coefficient_expansion}
\end{align}

\noindent where we have inserted $m-1$ complete set of scale invariant eigenstates, and defined:

\begin{equation}
|\psi_0 \rangle = \sum_l C_l(0) |l\rangle,
\end{equation}

\noindent with $C_l(0)$ being the initial expansion coefficients.

To make further progress, it is necessary to examine the time dependence of the deviation. For the cases under consideration, we write the deviation as:

\begin{equation}
\delta\tilde{H} (\tau) = \tilde{h} \left(\frac{\lambda(\tau)}{a}\right)^{\alpha},
\label{B:eq:def_delta_h}
\end{equation}

\noindent where $ \tilde{h}$ is an operator that contains the matrix structure of the deviation, $a$ is a length scale that characterizes the breaking of scale invariance, and $\alpha$ is the scaling of the deviation, see Appendix \ref{app:alpha}. The scaling factor, $\lambda(\tau)$, is given by:

\begin{align}
\lambda(\tau) &= \lambda_0 \sec(\tau), &  0&\leq  \tau < \pi/2.
\label{B:eq:lambda_tau}
\end{align}

\noindent The explicit forms of $\delta \tilde{H}(\tau)$ for the two-body problem and non-interacting quantum gases were calculated in Appendix \ref{app:derivation}.

The long time dynamics, $t \gg \lambda_0^2$, are determined by the behaviour of the deviation near $\tau = \pi/2$, which ultimately depends on $\alpha$. Motivated by the results of appendix \ref{app:alpha}, we consider the cases when: $\alpha \geq 1$, and $\alpha < 1$.  

\subsection{$\alpha \geq 1$}
For the case $\alpha \geq 1$, the deviation diverges as a function of time as:

\begin{equation}
\lambda^{\alpha}(\tau) \approx \frac{1}{(\pi/2 -\tau)^{\alpha}}.
\label{B:eq:lambda_large_tau}
\end{equation}

This implies that the dominant contribution to the expansion coefficients in the long time limit, $t \gg \lambda_0^2$, will be from $\tau \approx \pi/2$. In this limit, the time dependence from the exponential factors in Eq.~(\ref{B:eq:coefficient_expansion}) are irrelevant as they do not generate any divergences. For these terms, it is safe to set $\tau = \pi/2$. This results in the majority of the phase factors cancelling, reducing the calculation to evaluating:

\begin{align}
C_n^{(m)}(\tau) &\approx -i^m \left(\frac{\lambda_0}{a}\right)^{\alpha m}\sum_{l_m}\langle n| \tilde{h}^m | l_m \rangle e^{2i(n-l_m)} C_{l_m}(0) \nonumber \\
& \int_0^{\tau}d\tau_1 ...\int_0^{\tau_{m-1}}d\tau_m \frac{1}{(\pi/2-\tau_1)^{\alpha}} ...\frac{1}{(\pi/2-\tau_m)^{\alpha}}.
\label{B:eq:coefficient_expansion_second}
\end{align}

\noindent Eq.~(\ref{B:eq:coefficient_expansion_second}) is equivalent to a long time expansion of the coefficients, $C_n$, and is valid up to corrections of $\lambda_0^2/t$. If one were to evaluate Eq.~(\ref{B:eq:coefficient_expansion_second}) for perturbations with scalings, $0< \alpha <1$, one would find that the dominant contribution to the integrals is from short times. In this case, the dynamics will be similar to the case when $\alpha <0$, which will be discussed shortly.

The leading long time behaviour of the expansion coefficients, Eq.~(\ref{B:eq:coefficient_expansion_second}), for $\alpha \geq 1$, can be readily evaluated to give:

\begin{align}
\int_0^{\tau}d\tau_1 ...&\int_0^{\tau_{m-1}}d\tau_m \frac{1}{(\pi/2-\tau_1)^{\alpha}} ...\frac{1}{(\pi/2-\tau_m)^{\alpha}} \nonumber \\
&= \frac{1}{m!} \log^{m} \left(\frac{1}{\pi/2-\tau} \right), & \alpha = 1 \nonumber \\
&= \frac{1}{m!} \left(\frac{1}{\alpha-1} \frac{1}{(\pi/2-\tau)^{\alpha-1}}\right)^m, & \alpha >1
\end{align}

We now arrive at the final expression for the $mth$ order of the expansion coefficient:

\begin{align}
C_n&^{(m)}(\tau) \nonumber \\
&\approx \frac{1}{m!} \langle n | \left(i \left(\frac{\lambda_0}{a}\right) \tilde{V}_I \log(\pi/2 -\tau) \right)^m |\psi_0 \rangle, & \alpha&=1, \nonumber \\
& \approx \frac{1}{m!} \langle n | \left(-i \tilde{V}_I\frac{1}{\alpha -1} \left(\frac{\lambda_0}{a}\right)^{\alpha} \left(\frac{1}{\pi/2-\tau}\right)^{\alpha-1} \right)^m | \psi_0 \rangle, & \alpha &> 1,
\label{B:eq:final_mth_coefficient}
\end{align}

\noindent where we have defined:

\begin{equation}
\tilde{V}_I = e^{i \tilde{H} \pi/2} \tilde{h} e^{-i \tilde{H} \pi/2}.
\label{B:eq:V}
\end{equation}

With the aid of Eqs.~(\ref{B:eq:final_mth_coefficient}) and (\ref{B:eq:V}), it is possible to obtain the full expansion coefficient, valid in the large time limit, $t \gg \lambda_0^2$:

\begin{align}
C_n(\tau) &= \sum_m C_n^{(m)}(\tau) \nonumber \\
&\approx \langle n |\exp\left(i \frac{\lambda_0}{a} \tilde{V}_I \log(\pi/2-\tau)\right) | \psi_0 \rangle, & \alpha &=1, \nonumber \\
&\approx \langle n |\exp\left(-i  \frac{\tilde{V}_I}{\alpha-1} \left( \frac{\lambda_0}{a} \right)^{\alpha}\frac{1}{(\pi/2 - \tau)^{\alpha-1}}\right) | \psi_0 \rangle, & \alpha &>1.
\end{align}

\subsection{$\alpha <1$}

When $\alpha <1$, the deviation vanishes at large times, i.e. when $\tau$ tends to $\pi/2$. For sufficiently small deviations, i.e. for small scattering lengths $a \ll \lambda_0$, first order perturbation theory is sufficient at capturing the dynamics for all times:

\begin{align}
C_n(\tau) &= C_n(0) \nonumber \\
&- i \left(\frac{4 \pi a}{\lambda_0}\right)^{\alpha}\int_0^\tau d\tau' \cos^{\alpha}(\tau')\langle n | e^{i \tilde{H} \tau'} \tilde{h} e^{i \tilde{H} \tau'} | \psi_0 \rangle.
\label{B:eq:pt_weak}  
\end{align}

\noindent The expansion coefficients for $\alpha<1$ will have a well defined long time limit: $\lim_{\tau \rightarrow \pi/2}C_n(\tau) = C_n(\pi/2)$. For this reason, the wave function in the co-moving frame will eventually freeze. The only effect of the interactions will be to alter the profile of the wave function.

\section{Beat Amplitudes for Moment of Inertia and Contact}
\label{app:coefficients}

In the previous appendix, the dynamics of the wave function were evaluated near resonance. In this appendix we report the analytic expressions for the long time, $t\gg \lambda_0^2$, or equivalently, $\tau \approx \pi/2$, dynamics for the moment of inertia of an non-interacting quantum gas in the presence of an impurity, and for the contact in the two-body problem, both near resonance.

The moment of inertia for $N$-particles is defined as:

\begin{align}
\langle r^2 \rangle (t) &= \frac{1}{N}\sum_{i=1}^N r_i^2 = \frac{\lambda^2(t)}{N}\sum_{i=1}^N x_i^2 \nonumber \\
&= \lambda^2(t)\langle x^2 \rangle(\tau(t)),
\end{align}

\noindent where $\langle x^2 \rangle$ is the moment of inertia calculated in the co-moving frame. The long time dynamics of the moment of inertia, in the co-moving frame, has the following form:

\begin{align}
\lim_{t \rightarrow \infty}\langle x^2 \rangle(t)  &\approx A + B  \sin\left(v \frac{\lambda_0}{a} \log(t/\lambda_0^2) \right) \frac{\lambda_0^2}{t}\nonumber \\
&+ D  \sin^2\left(  \frac{v}{2} \ \frac{\lambda_0}{4 \pi a} \log(t/\lambda_0^2) \right).
\end{align}

\noindent The coefficients $A$, $B$, and $D$ are found by explicitly evaluating the expectation value. Here we quote the result:

\begin{align}
&A= \sum_{\lbrace n_i \rbrace, \lbrace m_i \rbrace =0}^{n_{max}} \sum_{P,Q} \frac{(\pm1)^{P+Q}}{N!} \psi(\lbrace m_i \rbrace) \psi(\lbrace n_i \rbrace) \nonumber \\
& \left[ \frac{1}{2N} \sum_{j=1}^N (4n_{Q_j}+1) \prod_{k=1,N} \delta_{m_{P_k},n_{Q_k}} \right. \nonumber \\
& \left. \frac{1}{N}\sum_{i=j}^N \sqrt{(2n_{Q_j}+1)(2n_{Q_j}+1)} \prod_{k\neq j = 1}^N \delta_{m_{P_k}, n_{Q_k}} \delta_{m_{P_j}, n_{Q_j}+1} \right]\label{C:eq:A}\\
&B= \sum_{\lbrace n_i \rbrace, \lbrace m_i \rbrace =0}^{N} \frac{(\pm 1)^{P+Q}}{N!} \frac{1}{2N} \sum_{j=1}^N \nonumber \\
& \left[ \sqrt{(2n_{Q_j}+1)(2n_{Q_j}+1)} \prod_{k\neq j = 1}^N \delta_{m_{P_k}, n_{Q_k}} \delta_{m_{P_j}, n_{Q_j}+1} \right. \nonumber \\
&\left. \left( \psi(\lbrace n_i \rbrace) \langle m_{P_1}, ... m_{P_N} | P_v | \psi_0 \rangle \right. \right. \nonumber \\
&\left. \left. - \psi(\lbrace m_i \rbrace) \langle n_{Q_1}, ... ,n_{Q_N} | P_v | \psi_0 \rangle \right)  \right]\label{C:eq:B} \\
&D= \sum_{\lbrace n_i \rbrace, \lbrace m_i \rbrace =0}^{N} \frac{(\pm 1)^{P+Q}}{N!} \frac{2}{N} \sum_{j=1}^N \left[ (4n_{Q_j}+1) \prod_{k=1,N} \delta_{m_{P_k},n_{Q_k}}  \right. \nonumber \\
& \left. \left( \langle n_{Q_1},..., n_{Q_N} | P_v | \psi_0 \rangle \langle m_{P_1},..., m_{P_N} | P_v | \psi_0 \rangle \right. \right. \nonumber \\
&\left. \left. - \langle m_{P_1},..., m_{P_N} | P_v | \psi_0 \rangle \psi(\lbrace n_i \rbrace)  \right. \right. \nonumber \\
&- \left. \left. \langle n_{Q_1},..., n_{Q_N} | P_v | \psi_0 \rangle \psi(\lbrace m_i \rbrace) \right) \right] \nonumber \\
& + \sum_{\lbrace n_i \rbrace, \lbrace m_i \rbrace =0}^{N} \frac{(\pm 1)^{P+Q}}{N!} \frac{2}{N} \sum_{j=1}^N \left[ \sqrt{(2n_{Q_j}+1)(2n_{Q_j}+1)} \right. \nonumber \\
&\left. \prod_{k\neq j = 1}^N \delta_{m_{P_k}, n_{Q_k}} \delta_{m_{P_j}, n_{Q_j}+1} \left[ \langle m_{P_1},..., m_{P_N} | P_v | \psi_0 \rangle \psi(\lbrace n_i \rbrace) \right. \right. \nonumber \\
& \left. \left. + \langle n_{Q_1},..., n_{Q_N} | P_v | \psi_0 \rangle \psi(\lbrace m_i \rbrace) \right. \right. \nonumber \\
&\left. \left. -\langle n_{Q_1},..., n_{Q_N} | P_v | \psi_0 \rangle \langle m_{P_1},..., m_{P_N} | P_v | \psi_0 \rangle \right] \right] \label{C:eq:D}
\end{align}

In Eqs.~(\ref{C:eq:A}), (\ref{C:eq:B}), and (\ref{C:eq:D}), $|\psi_0 \rangle$, is the fully symmetrized initial state:

\begin{align}
|\psi_0 \rangle &= \sum_{\lbrace n_i \rbrace} \left[ \psi({\lbrace n_i \rbrace}) \frac{1}{\sqrt{N!}}\sum_P (\pm 1)^P |n_{P_1}, ...,n_{P_N}\rangle \right]
\end{align}

The many body states are expanded in the single particle basis: $| n_1, ..., n_N \rangle$, where $n_i = 0,1, ..., n_{max}$, with $n_{max} = r_0^{-2} /\lambda_0^2$. The quantities $\psi(\lbrace n_i \rbrace)$ are the expansion coefficients for the many body states; they do not depend on the specific permutation of the indices, but rather the combination of indices. For this reason, the summations over the many body states is restricted to distinct combinations of single particle indices. The symmetry is fixed by summing over all the permutations of a given set of single particle indices, specified by the summation over $P$.

For the non-interacting quantum gas, the deviation from scale invariance, $\tilde{V}_I$ will only have a single non-zero eigenvalue of value $v$:

\begin{align}
v &= \sum_{n=0}^{n_{max}} f_n^2, & f_n &= \frac{\sqrt{2}}{\pi^{1/4}}\frac{(2n-1)!!}{\sqrt{(2n)!}}.
\label{C:eq:v}
\end{align}

\noindent For $N=1$, the state with eigenvalue $v$ is non-degenerate, however for multiple particle there are a number of degenerate states with eigenvalue, $v$. The operator, $P_v$, in Eqs.~(\ref{C:eq:A}), (\ref{C:eq:B}), and (\ref{C:eq:D}) is simply the projection operator onto all the eigenstates of the $N$-body deviation, $\tilde{V}_I$, with eigenvalue $v$.

Similarly, we quote the results for the contact of the relative motion in the interacting two-body problem. Near resonance, the asymptotic form of the contact is:

\begin{align}
\lim_{t \rightarrow \infty} C(t) \approx \frac{E}{\lambda(t)} + \frac{F}{\lambda(t)} \sin^2\left(\frac{v}{2} \ \frac{\lambda_0}{4 \pi a} \log\left(\frac{t}{\lambda_0^2}\right) \right). \nonumber \\
\end{align} 

\noindent The coefficients, $E$ and $F$, are given by:

\begin{align}
E &= \left| \sum_{n=0}^{n_{max}} f_n \frac{\sqrt{\pi}}{2} (-1)^n C_n(0) \right|^2, \label{C:eq:E} \\
F &= \left|\sum_{n=0}^{n_{max}} f_n \sqrt{\pi} (-1)^n \langle n | v \rangle \langle v | \psi_0 \rangle \right|^2 \nonumber \\
&- \sum_{n,n'=0}^{n_{max}} \frac{\pi}{2} f_n f_n' (-1)^{n-n'}  \nonumber \\
& \cdot \left(C_n'(0) \langle n | v \rangle \langle v | \psi_0 \rangle + C_n(0) \langle n' | v \rangle \langle v | \psi_0 \rangle \right).
\label{C:eq:F}
\end{align}

\noindent where $C_n(0)$ here are the expansion coefficients for the relative motion, $v$ and $f_n$ are defined in Eq.~(\ref{C:eq:v}), and:

\begin{equation}
\langle n | v \rangle = f_n/\sqrt{v}.
\end{equation}


\begin{thebibliography}{References}
\bibitem{Note} From this point on we neglect the spin coordinate. The inclusion of spin in the following study is straightforward.
\end{thebibliography}
\end{document}